\documentclass[epj]{svjour}
\usepackage{graphicx} 
\begin{document}
\title{Search for TeV Scale Physics in Heavy Flavour Decays}
\author{George W.S. Hou} 
%
%
\institute{Department of Physics, National Taiwan
 University, Taipei, Taiwan 10617, R.O.C.}
\date{Received: date / Revised version: date}
%
\abstract{
 The subject of heavy flavour decays as probes for physics at and
 beyond the TeV scale is covered from the experimental perspective.
 Emphasis is placed on the more traditional Beyond the Standard Model
 topics that have potential for impact in the early LHC era, and
 in anticipation of the B factory upgrade(s).
 The aim is to explain the physics, without getting too involved
 in the details, whether experimental or theoretical,
 to give the interested nonexpert a perspective on the Flavour/TeV link.
 We cover the forefront topics of $CP$ violation in $b\to s$ transitions
 involving penguin and box diagrams, and probes of
 charged Higgs, right-handed and scalar interactions.
 We touch briefly on $\Upsilon$ decay, $D^0$ mixing, rare $K$ decays,
 and lepton flavor violating probes in $\tau$ decay.
 Our own phemonenology work is often used for illustration.
\PACS{
      {PACS-key}{discribing text of that key}   \and
      {PACS-key}{discribing text of that key}
     } 
} 
\maketitle
\section{Introduction}
\label{intro}

As humans we aspire to reach up to the heavens, beyond the veiling
clouds of the v.e.v. scale. The conventional high energy approach,
such as the LHC, is like the fabled Jack climbing the bean stalk,
where impressions are that the Higgs boson may be just floating in
a low cloud close by. But then maybe not. It, or the something,
may lie up above the darker clouds of the v.e.v.! In this direct
ascent approach, Jack has to be fearful of the Giant, which in
this case could even be the projects like LHC and ILC themselves;
the cost is becoming so forbidding, Jack may not be able to
return. However, ``Jack" may not have to actually climb the bean
stalk: quantum physics allows him to stay on Earth, and let
virtual ``loops" do the work. The virtual Jack has no fear of
getting eaten by the Giant. This parable illustrates how flavour
physics offers probes of the TeV scale, at much reduced costs.
The flavour connection to TeV scale physics is typically through
loops.

A further ``parable" illustrates the potential for making impact.
Let us entertain a hypothetical ``What if?" question, by
forwarding to the recent past. On July 31, 2000, the BaBar
experiment announced at the Osaka conference the low value of
$\sin2\beta \sim 0.12$. The value for the equivalent $\sin2\phi_1$
from the Belle experiment was slightly higher, but also consistent
with zero. Within the same day, a theory paper appeared on the
arXiv~\cite{KN00}, entertaining the (New Physics) implications of
the low $\sin2\beta$ value. It seems some theorists have power to
``wormhole" into the future\footnote{
  This parable was meant as a joke, but as I was preparing for
  my SUSY2007 talk, the basis of this brief review, the paper
  ``Search for Future Influence from L.H.C''
   appeared~\cite{Nielsen}. So it was no joke after all.
 }$\,$!
A year later, however, both BaBar and Belle claimed the
observation of $\sin2\beta/\phi_1 \sim 1$, which turn out to be
consistent with Standard Model (SM) expectations. But, {\it what
if it stayed close to zero?} Well, it didn't. Otherwise, you would
have heard much more about it: a definite large deviation from the
SM has been found! For even in the last century, one expected from
indirect data that $\sin2\beta/\phi_1$ had to be nonzero within
SM.

Note that in SM, $\beta/\phi_1 = -\arg V_{td}$~\cite{PDG}. The
measurement of $\sin2\beta/\phi_1$ is the measurement of the $CP$
violating (CPV) phase in the $B^0$--$\bar B^0$ mixing matrix
element $M_{12}^d$. We recall that the discovery of $B^0$--$\bar
B^0$ mixing itself by the ARGUS experiment~\cite{ARGUS87} 20 years
ago was the first clear indication that $m_t$ is heavy, a decade
before the top quark was actually discovered at the Tevatron. With
the $B^0$--$\bar B^0$ mixing frequency $\Delta m_{B_d}$
proportional to $\vert V_{td}\vert^2\, m_t^2$, it is {\it the}
template of flavour loops as probes into high energy scales.
So let us learn from it.

\begin{figure}[b!]
\vskip-0.3cm\hskip0.2cm
\includegraphics[width=0.46\textwidth,height=0.23\textwidth,angle=0]{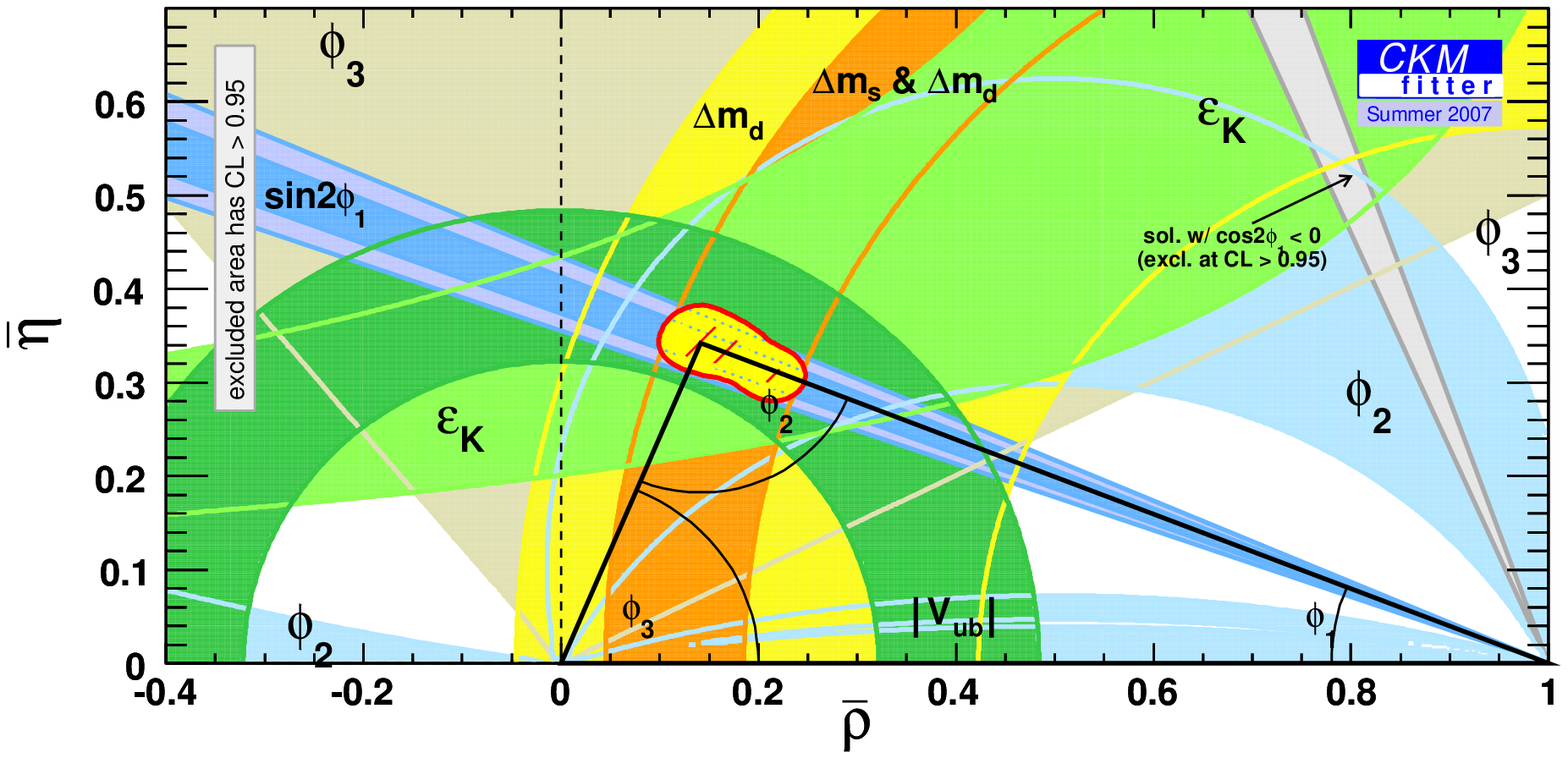}
\caption{
 CKM unitarity fit to all data as of summer 2007~\cite{HFAG}.
 }
 \label{fig:CKM07}       
\end{figure}

The $B^0$--$\bar B^0$ mixing amplitude $M_{12}^d$ is generated by
the box diagram involving two internal $W$ bosons and top quarks
in the loop. Normally, heavy particles such as the top would
decouple in the heavy $m_t \to \infty$ limit. However, the
longitudinal component of the $W$ boson, which is a charged Higgs
scalar that got eaten by the $W$ through spontaneous symmetry
breaking, couples to the top quark {\it mass}. This gives rise to
the {\it nondecoupling} of the top quark from the box diagram,
i.e. $M_{12}^d \propto (V_{tb}V_{td}^*)^2\, m_t^2$ to first
approximation. It illustrates the {\it Higgs affinity} of heavy
SM-like quarks (called chiral quarks), i.e. $\lambda_t \sim 1$,
which brings forth $V_{td}^{*2}$, the CPV phase of which is
$\sin2\beta/\phi_1$, which was measured by the B factories in
2001.

As we will only be interested in New Physics (NP), we note that
extensive studies at the B factories (and elsewhere) indicate that
$b\to d$ transitions are consistent with the SM~\cite{HFAG}, i.e.
no discrepancy is apparent with the CKM
(Cabibbo-Kobayashi-Maskawa) triangle~\cite{PDG}
\begin{eqnarray}
V_{ud}^*V_{ub} + V_{cd}^*V_{cb} + V_{td}^*V_{tb} = 0,
 \label{eq:btodTri}
\end{eqnarray}
which is the $db$ element of $V^\dag V = I$. This is illustrated
in Fig.~\ref{fig:CKM07}. An enormous amount of information and
effort has gone into this figure, the phase of $V_{td}^*V_{tb}$
being only one eminent entry. In general, we see no deviation from
CKM expectations.

What about $b\to s$ transitions? This will be our starting point
and the main theme.

The outline of this brief review is as follows.
In the next section we cover the main subject of CPV search in
loop-induced $b\to s$ transitions: the mixing-dependent CPV
difference $\Delta{\cal S}$ between $b\to c\bar cs$ and $s\bar qq$
processes; the direct CPV difference $\Delta{\cal A}_{K\pi}$
between $B^+$ and $B^0$; the status and prospects for measuring
the CPV phase $\sin2\Phi_{B_s}$ involving $B_s$ mixing, in
particular whether it could be large; and direct CPV in $B^+\to
J/\psi K^+$ decay.
In Sec.~\ref{sec:H+}, we turn to $b\to s\gamma$ and $B^+\to
\tau^+\nu$ to illustrate forefront probes of the charged Higgs
boson $H^+$. In Sec.~\ref{sec:EWP}, we use the forward-backward
asymmetry in $B\to K^*\ell^+\ell^-$ to illustrate how such
electroweak penguin processes probe the $bsZ$ vertex and related
physics, and $B\to K^{(*)}\nu\nu$ search as a window on light dark
matter. In Sec.~\ref{sec:RHscalar}, we use time-dependent CPV in
$B^0\to K_S\pi^0\gamma$ to illustrate the probes of right-handed
dynamics, and $B_s\to \mu^+\mu^-$ as probes of the extended Higgs
sector.
This brings us to a ``detour" in Sec.~\ref{sec:UpsNP}, to discuss
the utility of the bottomonium system as probes of light dark
matter and exotic light Higgs bosons.
We then turn briefly to loop effects in $D^0$ mixing and rare $K$
decays in Sec.~\ref{sec:D/K}, and lepton flavor violation in
$\tau$ decays in Sec.~\ref{sec:tau}, before closing with some
discussions, and offering our conclusion. As this is a
contribution to a memorial volume dedicated to Julius Wess, a
tribute is given as epilogue.
In an Appendix, we briefly introduce the mechanism for CPV.

\section{\boldmath CPV in $b \to s$: On Boxes and Penguins}
\label{btos}

The subject of CPV studies in charmless $b\to s$ transitions is
the current frontier of heavy flavour research. As $3\to 2$
transitions involving quarks, the subject also has $\tau\to \mu$
echoes in the lepton sector, which will be discussed later. We
focus on four topics: $\Delta{\cal S}$, $\Delta{\cal A}_{K\pi}$,
$\sin2\Phi_{B_s}$, and ${\cal A}_{B^+\to J/\psi K^+}$. Further
charmless $b\to s$ probes are discussed in subsequent sections.


\subsection{\boldmath The $\Delta{\cal S}$ Problem}
 \label{sec:DeltaS}

The B factories were built to measure time-dependent CPV (TCPV) in
the $B^0\to J/\psi K_S$ mode. One utilizes the coherent production
of $B^0\bar B^0$ pairs from $\Upsilon(4S)$ decay, and

\noindent \ \ 1) reconstruct one $B$ decay to a {\it $CP$
                 eigenstate},

\noindent \ \ 2) {\it tag} the other $B$ meson {\it flavour}
                 ($B^0$ or $\bar B^0$), and

\noindent \ \ 3) measure both the $B$ {\it decay vertices}.

\begin{figure}[b!]
\vskip-0.1cm \hskip0.6cm
\includegraphics[width=0.40\textwidth,height=0.28\textwidth,angle=0]{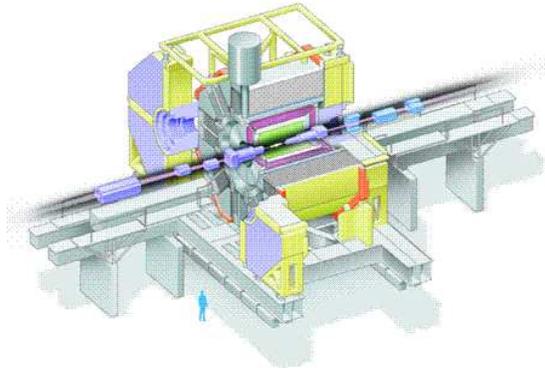}
\caption{Schematic picture of the Belle detector.}
 \label{fig:Belle}       
\end{figure}

\noindent The BaBar and Belle
 (the latter illustrated schematically in Fig.~\ref{fig:Belle})
detectors are rather similar, differing basically only in the
particle identification detector (PID) used for flavour tagging,
the task of charged $K/\pi$ separation at various energies. Belle
uses Aerogel Cherenkov Counters, a threshold device, and a TOF
system, while BaBar uses the the DIRC, basically a system of
quartz bars that guide Cherenkov photons and project them into a
water tank at the back end of the detector.

The real novelty of the B factories is the asymmetric beam
energies. The $\beta\gamma$ factor for the produced $\Upsilon(4S)$
is 0.56 and 0.43, respectively. Boosting the $B^0$ and $\bar B^0$
mesons allow the time difference $\Delta t \cong \Delta
z/\beta\gamma c$ to be inferred from the decay vertex difference
$\Delta z$ in the boost direction, while the proximity of
$2m_{B^0}$ to $m_{\Upsilon(4S)}$ means rather minimal lateral
motion. Both the PEP-II and KEKB accelerators were commissioned in
1999 with a roaring start. By 2001, KEKB outran PEP-II in the
instantaneous luminosity, and surpassing in integrated luminosity
as well in the following year. In April 2008, PEP-II dumped its
beam for the last time.

With the measurement of TCPV in $B^0\to J/\psi K_S$ settled in
summer 2001, attention quickly turned to the $b\to s$ penguin
modes, where a virtual gluon is emitted from the virtual top quark
in the vertex loop. Let us take $B^0 \to \phi K_S$ as example,
where the virtual gluon pops out an $s\bar s$ pair. The $b\to s$
penguin amplitude is practically real within SM, just like the
tree level $B^0\to J/\psi K_S$. This is because $V_{us}^*V_{ub}$
is very suppressed. Thus, SM predicts
\begin{eqnarray}
{\cal S}_{\phi K_S} \cong \sin2\phi_1/\beta,\ \ \ \ \ \ {\rm (SM)}
 \label{eq:SphiK}
\end{eqnarray}
where ${\cal S}_{\phi K_S}$ is the analogous TCPV measure in the
$B^0 \to \phi K_S$ mode. New physics induced flavour changing
neutral current (FCNC) and CPV effects, such as having
supersymmetric (SUSY) particles in the loop (for example, $\tilde
b_R$-$\tilde s_R$ squark mixing) could break this equality,
prompting the experiments to search vigorously.

\begin{figure}[t!]
\vskip-0.2cm \hskip1.2cm
\includegraphics[width=0.33\textwidth,height=0.45\textwidth,angle=0]{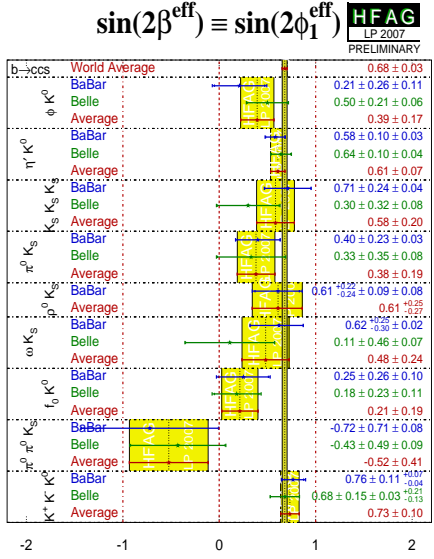}
\caption{
 Measurements of ${\cal S}_f$ in $b\to s$ penguin modes~\cite{HFAG}.
 See footnote 2 for comment on the $B^0\to f_0(980)K_S$ mode.}
 \label{fig:Ssqq07}       
\end{figure}

Many might remember the big splash made by Belle in summer 2003,
where ${\cal S}_{\phi K_S}$ was found to be opposite in
sign~\cite{phiKs_Belle03} to $\sin2\phi_1/\beta$, where the
significance of deviation was more than 3$\sigma$. But the
situation softened by 2004 and it is now far less dramatic. But
some deviation has persisted in an interesting if not nagging kind
of way. Comparing to the average of ${\cal S}_{c\bar cs} = 0.681
\pm 0.025$~\cite{HFAG} over $b\to c\bar cs$ transitions, ${\cal
S}_f$ is smaller in practically all $b\to s\bar qq$ modes measured
so far (see Fig.~\ref{fig:Ssqq07}), with the naive mean\footnote{
 We use the LP2007 update that excludes the new
 ${\cal S}_{f_0(980)K_S}$ result from BaBar.
 The Heavy Flavour Averaging Group (HFAG) itself
 warns ``treat with extreme caution" when using this BaBar result~\cite{HFAG}.
 The value is larger than ${\cal S}_{c\bar cs}$ and is very
 precise, with errors 3 times smaller than the $\phi K_S$ mode,
 but $f_0(980)K_S$ actually has smaller branching ratio!
 The BaBar result needs confirmation from Belle.
 }
of ${\cal S}_{s\bar qq} = 0.56 \pm 0.05$~\cite{HFAG}. The
deviation is only 2.2$\sigma$, and the significance has been
slowly diminishing. We stress, however, that the persistence over
several years, and in multiple modes, together make this
``$\Delta{\cal S}$ problem" a potential indication for New Physics
from the B factories, and should be taken seriously.

The point is that theoretical studies, although troubled by
hadronic effects, all give ${\cal S}_{s\bar qq}$ values that are
{\it above} ${\cal S}_{c\bar cs}$. A model-independent geometric
approach suggests~\cite{SMH} that, with enough precision, a
deviation as little as a couple of degrees would indicate New
Physics. Alas, BaBar has ended its data taking, while Belle would
stop for (hopeful) upgrade after reaching 1 ab$^{-1}$, so the
dataset for analysis can only double within the present B factory
era, which is coming to an end. One may think that the LHC, which
is turning on in 2008, the LHCb experiment in particular, could
make great impact. But because of lack of good vertices, or
presence of neutral ($\pi^0$, $\gamma$) particles, in the leading
channels of $\eta^\prime K_S$, $\phi K_S$ and $K_S \pi^0$, the
situation may not improve greatly with LHCb data. Thus, {\it the
$\Delta{\cal S}$ problem would need a Super B Factory to clarify}.

\subsection{\boldmath The $\Delta{{\cal A}_{K\pi}}$ Problem}
\label{sec:DeltaAKpi}

There is a second possible indication for physics beyond SM (BSM)
in $b\to s\bar qq$ decays. It is by now widely known because of
Belle effort, and, unlike the $\Delta{\cal S}$ situation,
experimentally it is very firm.

Just 3 years after the observation of TCPV in $B^0\to J/\psi K^0$,
direct CPV (DCPV) in the B system was claimed in 2004 between
BaBar and Belle~\cite{PDG}. This attests to the prowess of the B
factories, as it took 35 years for the same evolution in the K
system. The CDF experiment recently joined the club, with results
consistent with the B factories. The current world
average~\cite{HFAG} is ${\cal A}_{K^+\pi^-} = -9.7 \pm 1.2\ \%$.
This by itself does not suggest New Physics, but rather, it
indicates the presence of a finite strong phase between the strong
penguin (P) and tree (T) amplitudes, where the latter provides the
weak phase (for a primer on CPV, see the Appendix). Most QCD based
factorization approaches failed to predict ${\cal A}_{K^+\pi^-}$.

Even in 2004, however, there was a whiff of a
puzzle~\cite{belle04}. In contrast to the negative value for $B^0
\to K^+\pi^-$, DCPV in the charged $B^+\to K^+\pi^0$ mode was
found to be consistent with zero for both Belle and BaBar. The
difference between the charged and neutral mode has steadily
strengthened, where the current~\cite{HFAG} ${\cal A}_{K^+\pi^0} =
+5.0 \pm 2.5\ \%$ shows some significance for the sign being
positive.
In a recent paper published in {\it Nature}, the Belle
collaboration used 535M $B\bar B$ pairs to demonstrate a
difference~\cite{BelleNature}
\begin{eqnarray}
\Delta{{\cal A}_{K\pi}}
 \equiv {\cal A}_{K^+\pi^0} - {\cal A}_{K^+\pi^-}
 = +0.164 \pm 0.037,
 \label{eq:DeltaAKpiBelle}
\end{eqnarray}
with 4.4$\sigma$ significance by a single experiment, and
emphasized the possible indication for New Physics. The world
average~\cite{HFAG},
\begin{eqnarray}
\Delta{{\cal A}_{K\pi}} = 0.147 \pm 0.027,
 \label{eq:DeltaAKpi}
\end{eqnarray}
is now beyond $5\sigma$.

\begin{figure}[t!]
\vskip0.3cm\hskip0.3cm
\includegraphics[width=0.45\textwidth,height=0.31\textwidth,angle=0]{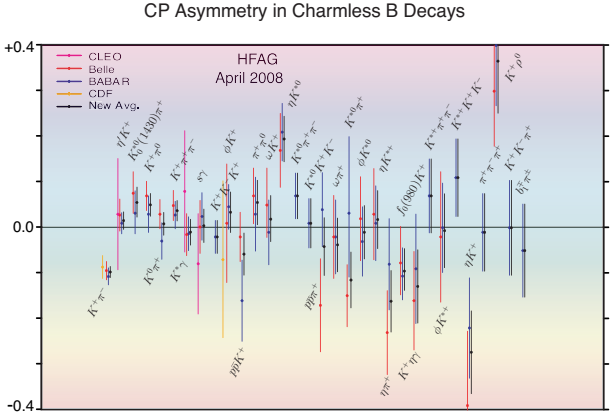}
\vskip0.3cm
 \caption{
 HFAG plot for DCPV measurements.
 The difference between ${\cal A}_{K^+\pi^0}$
 and ${\cal A}_{K^+\pi^-}$ could indicate New Physics.}
 \label{fig:ACP}       
\end{figure}

We plot in Fig.~\ref{fig:ACP} the current status of DCPV in $B$
decays. ${\cal A}_{K^+\pi^-}$ is clearly established, but no other
mode reaches the similar level of significance, and there is a
wide scatter in central values. So why is the $\Delta{{\cal
A}_{K\pi}}$ difference a puzzle, that it might indicate New
Physics?

For the $B^0$ decay mode, one has
\begin{eqnarray}
{\cal M}(B^0 \to K^+\pi^-)
 \propto T + P \propto r \, e^{i\phi_3} + e^{i\delta},
 \label{eq:MKpi}
\end{eqnarray}
where $\phi_3 = \arg V_{ub}^*$, $\delta$ is the strong phase
difference between the tree amplitude $T$ and strong penguin
amplitude $P$, and $r \equiv \vert T/P\vert$. It is the
interference between the two kinds of phases (Appendix A) that
gives rise to DCPV, i.e. ${\cal A}_{K^+\pi^-} \equiv {\cal A}_{\rm
CP}(K^+\pi^-)$.

Note that for TCPV, $\delta = \Delta m_B \Delta t$, where $\Delta
m_B$ is the already well measured $B^0$-$\bar B^0$ oscillation
frequency, and $\Delta t$ is part of the time-dependent
measurement. This is the beauty~\cite{BigiSanda} of mixing
dependent CPV studies, that is is much less susceptible to
hadronic effects, especially in single amplitude tree dominant
processes such as $B^0\to J/\psi K^0$. One has {\it direct access}
to the CPV phase of the $B^0$-$\bar B^0$ mixing amplitude.
In comparison, DCPV relies on the presence of strong interaction
phase differences. The hadronic nature of these $CP$ invariant
phases make them difficult to predict. Although DCPV is one of the
simplest things to measure (counting experiment), the strong phase
difference in a decay amplitude is usually hard to extract
experimentally.


%

The $B^+\to K^+\pi^0$ decay amplitude is similar to the $B^0\to
K^+\pi^-$ one, up to subleading corrections,
\begin{eqnarray}
\sqrt{2}{\cal M}_{K^+\pi^0} - {\cal M}_{K^+\pi^-}
 \propto C + P_{\rm EW},
 \label{eq:DeltaMKpi}
\end{eqnarray}
where $C$ is the colour-suppressed tree amplitude, while $P_{\rm
EW}$ is the electroweak penguin (replacing the virtual gluon in
$P$ by $Z$ or $\gamma$) amplitude. In the limit that these
subleading terms vanish, one expects $\Delta{{\cal A}_{K\pi}} \sim
0$, which was broadly expected to be the case, but contrary to the
experimental result of Eq.~(\ref{eq:DeltaAKpi}).

Could $C$ be greatly enhanced? This is the attitude taken by
many~\cite{Gronau}. Indeed, fitting with data, one finds $\vert
C/T\vert > 1$ is needed~\cite{BL07}, in contrast to the very tiny
suggested value of 10 years ago~\cite{Neubert98}. Furthermore, as
the amplitude $C$ has the same weak phase $\phi_3$ as $T$, the
enhancement of $C$ has to contrive in its strong phase structure,
to cancel the effect of the strong phase difference $\delta$
between $T$ and $P$ that helped induce the sizable ${\cal
A}_{K^+\pi^-}$ in the first place. The amount of finesse needed is
therefore considerable.

It should be noted that this difference was not anticipated in any
calculations beforehand. In perturbative QCD factorization (PQCD)
calculations at next to leading order~\cite{NLOPQCD}, taking cue
from data (i.e. after the experimental fact), $C$ does move in the
right direction, but insufficiently so. For QCD factorization
(QCDF), it has been declared~\cite{Beneke} that $\Delta {\cal
A}_{CP}$ is difficult to explain, that it would need very large
and {\it imaginary} $C$ (or electroweak penguin), which is ``Not
possible in SM plus factorization [approach]." In the rather
sophisticated Soft Colinear Effective Theory (SCET)
approach~\cite{SCET}, ${\cal A}_{K^+\pi^0}$ is actually predicted
to be even more negative than ${\cal A}_{K^+\pi^-}$, where the
latter has been taken as input. But this is a problem for SCET
itself, rather than with experiment.

The other option is to have a large CPV contribution from the
electroweak penguin~\cite{BL07,HNS05,Peskin}. The interesting
point is that {\it this calls for a New Physics CPV phase}, as it
is known that $P_{\rm EW}$ carries practically no weak phase
within SM, and has almost the same strong phase as
$T$~\cite{NR98}. So what NP can this be? Note that this would not
so easily arise from SUSY, since SUSY effects tend to be of the
``decoupling" kind, compared to the {\it nondecoupling} of the top
quark effect already present in the $Z$ penguin loop. The latter
is very analogous to what happens in box diagrams.

So, can there be more {\it nondecoupled} quarks beyond the top in
the $Z$ penguin loop? This is the sequential fourth generation,
which would naturally bring into $b\to s$ {\it electroweak
penguin} $P_{\rm EW}$ (but not so much in the strong penguin $P$)
a new CPV phase, in the new CKM product $V_{t's}^*V_{t'b}$. It was
shown~\cite{HNS05} that Eq.~(\ref{eq:DeltaAKpiBelle}) can be
accounted for in this extension of SM. We will look further into
this, after we discuss NP prospects in $B_s$ mixing.

With the two hints for NP in $b\to s$ penguin modes, i.e.
$\Delta{\cal S}$ (TCPV) and $\Delta{{\cal A}_{K\pi}}$ (DCPV), one
might expect possible NP in $B_s$ mixing. Note that recent results
for $\Delta m_{B_s}$ and $\Delta \Gamma_{B_s}$ are SM-like, but
the real test clearly should be in the CPV measurables
$\sin2\Phi_{B_s}$ and $\cos2\Phi_{B_s}$, as the NP hints all
involve CPV.

\subsection{\boldmath $B_s$ Mixing and $\sin2\Phi_{B_s}$}
\label{sec:sin2PhiBs}

The oscillation between $B_s^0$ and $\bar B_s^0$ mesons is too
fast for B factories. This brings us to the hadronic collider
environment, which enjoys a large boost for produced $B$ mesons.
After a slow start of the Tevatron Run II, the CDF and
D$\emptyset$ experiments have recently reached $\sim 4$ fb$^{-1}$
integrated luminosity per experiment, and expect to accumulate an
overall of 6-8 fb$^{-1}$ per experiment throughout the Tevatron
Run II lifetime.

Despite the earlier announcement made by D$\emptyset$ in Winter
2006, the CDF experiment had the advantage of a special two-track
trigger. By Summer 2006, based on 1 fb$^{-1}$ data, $B_s$ mixing
became a precision measurement~\cite{Bsmix_CDF06},
\begin{eqnarray}
 \Delta m_{B_s} = 17.77 \pm 0.10 \pm 0.07\ {\rm ps}^{-1}.
 \label{eq:DeltamBs}
\end{eqnarray}

We remark that, if one takes the current nominal values for
$f_{B_s}$ e.g. from lattice studies, {\it the result of
Eq.~(\ref{eq:DeltamBs}) seems a bit on the small side}. Recall
that, before the experimental measurement precipitated, fitting to
data and information other than $\Delta m_{B_s}$ itself, the
fitted values from the CKMfitter and UTfit groups tended to be
larger than 20 ps$^{-1}$. The situation may be even more serious.
CLEO~\cite{fDsCLEO07} and Belle~\cite{fDsBelle08} have measured
$f_{D_s}$ by measuring $D_s^+ \to \ell^+\nu$ decays, and the
measured values are considerably higher than current lattice
results. If this carries over to $f_{B_s}$, the SM expectation for
$\Delta m_{B_s}$ would definitely be above 20 ps$^{-1}$, and one
may need some ``New Physics" to bring it down to the level of
Eq.~(\ref{eq:DeltamBs}). Unfortunately, because of the large
hadronic uncertainties in $f_{B_s}^2B_{B_s}$, one cannot take this
as a hint for New Physics. One has to turn to CPV that is less
prone to hadronic physics.

Analogous to the case for $B_d$ oscillations, the amplitude for
$B_s$ mixing in SM behaves as $M_{12}^s \propto (V_{tb}V_{ts}^*)^2
\, m_t^2$ to first approximation, and CPV in $B_s$ mixing is
controlled by the phase of $V_{ts}$. Since $\vert
V_{us}^*V_{ub}\vert$ is rather small, unlike the analogous
Eq.~(\ref{eq:btodTri}) for $b\to d$ transitions, the triangle
relation
\begin{eqnarray}
V_{us}^*V_{ub} + V_{cs}^*V_{cb} + V_{ts}^*V_{tb} = 0
 \label{eq:btosTri} \\
\ \;
 \Longrightarrow \ \;
V_{ts}^*V_{tb} \simeq - V_{cb},
 \label{eq:VtsVtbSM}
\end{eqnarray}
i.e. collapsing to approximately a line, and $V_{ts}^*V_{tb}$ is
practically real (in the standard phase convention~\cite{PDG} that
$V_{cb}$ is real). In practice, $\Phi_{B_s} \equiv -\arg V_{ts}
\sim -0.02$~rad in SM (the actual definition being $2\Phi_{B_s} =
\arg M_{12}^s$, if one ignores absorptive parts), is tiny compared
to $\Phi_{B_d} \equiv -\arg V_{td} = \beta/\phi_1 \sim 0.37$~rad.
With $\Phi_{B_s}$ at the percent level, only the LHCb experiment,
which is designed for $B$ physics studies at the LHC, would have
enough sensitivity to probe it. Thus, it is well known that
$\sin2\Phi_{B_s}$, the analogue of $\sin2\phi_1/\beta$ for $B_d$,
is an excellent window on BSM. That is, {\it any} observation that
deviates from
\begin{eqnarray}
\sin2\Phi_{B_s}\vert^{\rm SM} \cong -0.04,
 \label{eq:sin2PhiBsSM}
\end{eqnarray}
would be indication for New Physics. In SUSY, this could be
squark-gluino loops with $\tilde s$-$\tilde b$ mixing.

\begin{figure}[t!]
\vskip0.3cm\hskip1.2cm
\includegraphics[width=0.35\textwidth,height=0.24\textwidth,angle=0]{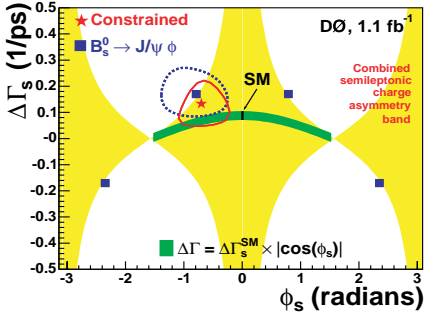}
\caption{
 Combined analysis of $A_{SL}$, $A_{SL}^s$ and lifetime
 difference in $B_s \to J/\psi \phi$ by D$\emptyset$~\cite{cos2PhiBs_Dzero07},
 based on 1.1 fb$^{-1}$ data.}
 \label{fig:phisDzero}       
\end{figure}

\subsubsection{$\Delta\Gamma_{B_s}$ approach to $\phi_{B_s}$}

Let us first briefly comment on the approach through width mixing,
i.e. $\Delta\Gamma_{B_s}$ and $\phi_{B_s}$ from untagged $B_s^0
\to J/\psi \phi$ and other lifetime studies. Here, the
D$\emptyset$ experiment has made a concerted effort on dimuon
charge asymmetry $A_{SL}$, the untagged single muon charge
asymmetry $A_{SL}^s$, and the lifetime difference in untagged $B_s
\to J/\psi \phi$ decay (hence does not involve oscillations),
using a dataset of 1.1 fb$^{-1}$. D$\emptyset$ holds the advantage
in periodically flipping magnet polarity to reduce the systematic
error on $A_{SL}$. Combining the three studies, they probe the CPV
phase $\cos2\Phi_{B_s}$ via
\begin{eqnarray}
\Delta \Gamma_{B_s} = \Delta \Gamma_{B_s}^{\rm CP}\,
\cos2\Phi_{B_s},
 \label{eq:cos2PhiBs}
\end{eqnarray}
where $\Delta \Gamma_{B_s}^{\rm CP}\cong \Delta \Gamma_{B_s}^{\rm
SM}$. The main result of interest is given in
Fig.~\ref{fig:phisDzero}, where $\phi_s = \Phi_{B_s}$ and
$\Delta\Gamma_s = \Delta\Gamma_{B_s}^{\rm CP}$. The fitted width
difference of $0.13\pm 0.09$ ps$^{-1}$ is still larger than the SM
expectation~\cite{LN07} of
\begin{eqnarray}
\Delta \Gamma_{B_s}\vert^{\rm SM} = 0.096\pm 0.039
                                   {\rm \ ps}^{-1},
 \label{eq:DGsSM}
\end{eqnarray}
but certainly not inconsistent. The extracted ``first" measurement
of $\vert\Phi_{B_s}\vert = 0.70^{+0.39}_{-0.47}$ is slightly off
zero. Given the large errors, it is both consistent with SM
expectation but certainly allows for NP. The details can be found
in Ref.~\cite{cos2PhiBs_Dzero07}, which is somewhat technical.
For a phenomenological digest, see Ref.~\cite{HM07}.
A more recent CDF study~\cite{DGs_CDF08} of $B_s \to J/\psi \phi$
using 1.7 fb$^{-1}$ data finds $\Delta \Gamma_{B_s} =
0.076^{+0.059}_{-0.063}\pm0.006$ ps$^{-1}$, assuming $CP$
conservation, which is consistent with the SM expectation of
Eq.~(\ref{eq:DGsSM}). Overall, our comment is that the
$\cos2\Phi_{B_s}$ approach is somewhat a ``blunt instrument".

\subsubsection{Prospects for $\sin2\Phi_{B_s}$ Measurement}

The more direct approach to measuring $\sin2\Phi_{B_s}$ is via
tagged TCPV study of $B_s \to J/\psi \phi$. Let us focus on the
shorter term prospects.

$B_s \to J/\psi \phi$ decay is analogous to $B_d \to J/\psi K_s$,
except it is a $VV$ final state. Thus, besides measuring the decay
vertices, one also needs to perform an angular analysis to
separate the $CP$ even and odd components.
As $J/\psi$ is reconstructed in say the dimuon final state, CDF
and D$\emptyset$ should have comparable sensitivity. Assuming 8
fb$^{-1}$ per experiment (which may be questionable), the Tevatron
could reach an ultimate sensitivity of
\begin{eqnarray}
\sigma(\sin2\Phi_{B_s}) \sim 0.2/\sqrt{2}.\ \ \
 {\rm (Tevatron\ combined)}
 \label{eq:sin2PhiBs_TeV}
\end{eqnarray}
However, just now the LHC magnets are cooling down towards
running. How fast can LHC turn on and produce physics results? We
will have to wait and see, but some training period is expected. I
will adopt a conservative estimate~\cite{Nakada07} for the ``first
year" (a floating concept in actual calendar terms) running of
LHC: 2.5 fb$^{-1}$ for ATLAS and CMS, and 0.5 fb$^{-1}$ for LHCb.
Assuming this, the projection for ATLAS is
$\sigma(\sin2\Phi_{B_s}) \sim 0.16$, not better than the Tevatron,
while for LHCb one has $\sigma(\sin2\Phi_{B_s}) \sim 0.04$. While
the situation would be rather volatile, these sensitivities,
listed side by side in Table 1, can be viewed as reference values
for 2009, perhaps even Winter conferences for 2010 or beyond.

%
\begin{table}[t!] \caption{Rough sensitivity to
$\sin2\Phi_{B_s}$ ca. 2009.}
\label{tab:sigma_SBs}       
\begin{center}
\begin{tabular}{cccc}
\hline\noalign{\smallskip}
& CDF/D{$\emptyset$} & ATLAS/CMS & LHCb   \\
\noalign{\smallskip}\hline\noalign{\smallskip}
$\sigma(\sin2\Phi_{B_s})$ & 0.2/expt & 0.16/expt & 0.04 \\
$\int {\cal L}dt$ & (8 fb$^{-1}$) & (2.5 fb$^{-1}$) & (0.5 fb$^{-1}$) \\
\noalign{\smallskip}\hline
\end{tabular}
\end{center}
\end{table}

\begin{figure}[t!]
\vskip0.3cm \hskip0.9cm
\includegraphics[width=0.40\textwidth,height=0.23\textwidth,angle=0]{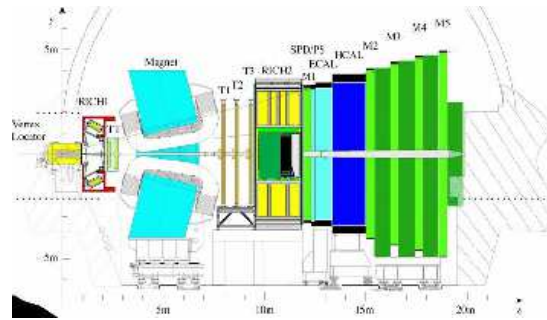}
\vskip0.35cm
 \caption{
 The LHCb detector.}
 \label{fig:LHCb}       
\end{figure}

If SM again holds sway, LHCb would clearly be the winner, since
$\sigma(\sin2\Phi_{B_s}) \sim 0.04$ starts to probe the SM
expectation of Eq.~(\ref{eq:sin2PhiBsSM}). This is not surprising,
as the LHCb detector (see Fig.~\ref{fig:LHCb}) has a forward
design for the purpose of $B$ physics. It takes advantage of the
large collider cross section for $b\bar b$ production, while
implementing a fixed-target-like detector configuration that
allows more space for devices such as RICH detectors for PID.
We wish to stress, however, that {\it 2009 looks rather
interesting}
--- Tevatron could get really lucky: it could glimpse the value of
$\sin2\Phi_{B_s}$ {\it only if} its strength is large; but {\it if
$\vert \sin2\Phi_{B_s}\vert$ is large, it would definitely
indicate New Physics}. Thus, {\bf the Tevatron has the chance to
preempt LHCb and carry away the glory of discovering physics
beyond the Standard Model in \boldmath{$\sin2\Phi_{B_s}$}}
(publicly stressed since early 2007).
Maybe the Tevatron should even run longer, especially if LHC
dangles. This can add to the existing competition on Higgs search
between Tevatron and the LHC.

\subsubsection{\bf\boldmath
              \ Can $\vert\sin2\Phi_{B_s}\vert > 0.5$?}

The answer should clearly be in the positive. We provide some
phenomenological insight as an existence proof, at the same time
attempting to link with the hints for New Physics discussed in the
two previous subsections. That is, it is of interest to explore
whether the $\Delta B = 1$ $b\to s$ processes of
Secs.~\ref{sec:DeltaS} and \ref{sec:DeltaAKpi} have implications
for the $\Delta B = 2$ $b\bar s \to s\bar b$ process.

One can of course resort to squark-gluino box diagrams. Note,
however, that squark-gluino loops, while possibly generating
$\Delta{\cal S}$, cannot really move $\Delta {\cal A}_{K\pi}$
because their effects are decoupled in $P_{\rm EW}$. If one wishes
to have contact with both hints for NP in $b\to s$ transitions
from the B factories, then one should pay attention to some common
nature between $b\to s$ electroweak penguin and the $B_s$ mixing
box diagram. If there are new {\it nondecoupled} quarks in the
loop, then both $\Delta {\cal A}_{K\pi}$ and $\Delta{\cal S}$
could be touched. It also affects $B_s$ mixing, as it is well
known that the top quark effect in electroweak penguin and box
diagrams are rather similar. Such new nondecoupled quarks are
traditionally called the 4th generation quarks $t^\prime$ and
$b^\prime$. The $t^\prime$ quark in the loop adds a term
$V_{t's}^*V_{t'b} \equiv r_{sb}\, e^{i\phi_{sb}}$ to
Eq.~(\ref{eq:btosTri}), bringing in the additional NP CPV phase
$\arg(V_{t's}^*V_{t'b}) \equiv \phi_{sb}$ with even larger {\it
Higgs affinity}, $\lambda_{t^\prime}
> \lambda_t \simeq 1$.
Dynamically speaking, this is no different from the SM.

It was shown~\cite{HNS05} that the 4th generation could account
for $\Delta {\cal A}_{K\pi}$, and $\Delta {\cal S}$ then moves in
the right direction~\cite{HLMN07}. This was done in the PQCD
approach up to next-to-leading order (NLO), which is the state of
the art. We note that PQCD is the only QCD-based factorization
approach that {\it predicted}~\cite{PQCD03} both the strength and
sign of ${\cal A}_{\rm CP}(B^0\to K^+\pi^-)$. At NLO, the $\Delta
{\cal A}_{K\pi}$ saw improvement by enhancement of
$C$~\cite{NLOPQCD}, although this also demonstrated that a
calculation approach could not generate $\vert C/T\vert > 1$. It
is nontrivial, then, that incorporating the nondecoupled 4th
generation $t^\prime$ quark to account for $\Delta {\cal
A}_{K\pi}$, can also move $\Delta {\cal S}$ in the right
direction.

The really exciting implication, however, is the impact on
$\sin2\Phi_{B_s}$: the $t^\prime$ effect in the box diagram also
enjoys nondecoupling. As the difference of $\Delta {\cal
A}_{K\pi}$ in Eq.~(\ref{eq:DeltaAKpi}) is large, both the strength
and phase of $V_{t's}^*V_{t'b}$ is sizable~\cite{HNS05}, and the
phase is not far from maximal. As we have mentioned, a near
maximal phase from $t^\prime$ allows precisely the minimal impact
on $\Delta m_{B_s}$, as it adds only in quadrature to the real
contribution from top. But it makes the maximal impact on
$\sin2\Phi_{B_s}$. Furthermore, the $t^\prime$ effect can
partially cancel against too large a $t$ contribution in the real
part, if the indication for large $f_{D_S}$ from experiment is
carried over to a larger $f_{B_s}$ value than current lattice
results.

\begin{figure}[t!]
\vskip0.2cm\hskip0.2cm
\includegraphics[width=0.23\textwidth,height=0.155\textwidth,angle=0]{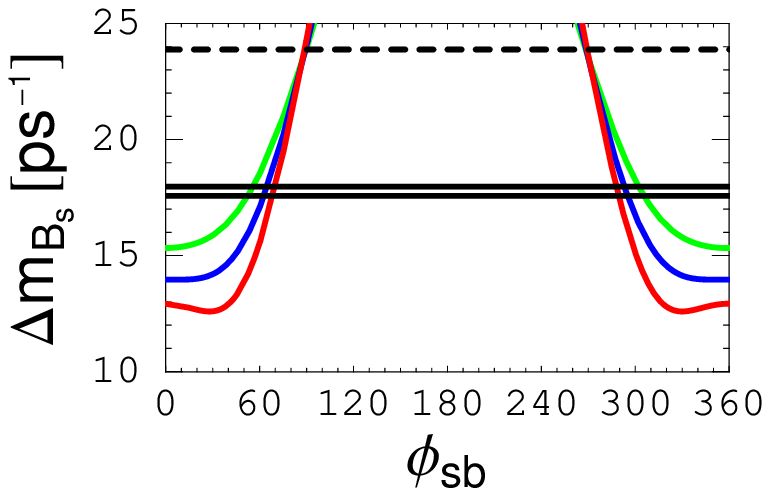}
\vskip-2.8cm\hskip4.3cm
\includegraphics[width=0.23\textwidth,height=0.155\textwidth,angle=0]{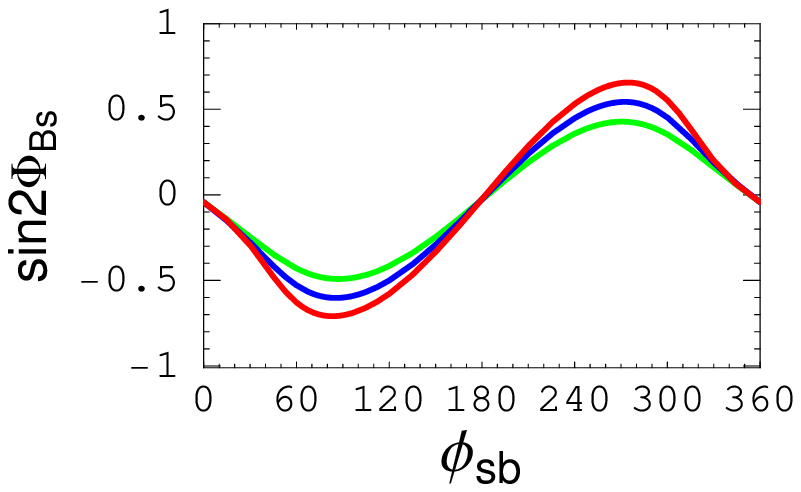}
\vskip0.35cm
 \caption{
 $\Delta m_{B_s}$ and $\sin2\Phi_{B_s}$ vs $\phi_{sb} \equiv \arg V_{t's}^*V_{t'b}$
 for 4th generation extension of SM~\cite{HNSprd},
 for $\vert V_{t's}^*V_{t'b}\vert = 0.02$, 0.025, 0.03.
 The dashed horizontal line is the nominal 3 generation SM expectation
 taking $f_{B_s}\sqrt{B_{B_s}} = 295$ MeV,
 and the solid band is the experimental measurement by CDF~\cite{Bsmix_CDF06}.
 The narrow range implied by $\Delta m_{B_s}$ measurement project
 out large values for $\sin2\Phi_{B_s}$, where the right branch is excluded
 by the sign of $\Delta {\cal A}_{K\pi}$, Eq.~(\ref{eq:DeltaAKpi}).
 }
 \label{fig:sin2PhiBs4th}       
\end{figure}

We show in Fig.~\ref{fig:sin2PhiBs4th} the variation of $\Delta
m_{B_s}$ and $\sin2\Phi_{B_s}$ with respect to the new CPV phase
$\phi_{sb} \equiv \arg V_{t's}^*V_{t'b}$ in the 4th generation
model, for the nominal $m_{t^\prime} = 300$ GeV and $r_{sb} \equiv
\vert V_{t's}^*V_{t'b} \vert =$ 0.02, 0.025, and 0.03, where
stronger $r_{sb}$ gives larger variation. Using the central value
of $f_{B_s}\sqrt{B_{B_s}} = 295 \pm 32$ MeV, we get a nominal 3
generation value of $\Delta m_{B_s}\vert^{\rm SM} \sim 24$
ps$^{-1}$, which is the dashed line. The CDF measurement of
Eq.~(\ref{eq:DeltamBs}) is the rather narrow solid band, attesting
to the precision already reached by experiment, and that it is
below the dashed line.
Combining the information from $\Delta {\cal A}_{K\pi}$, $\Delta
m_{B_s}$ and ${\cal B}(b\to s\ell^+\ell^-)$, the predicted value
is~\cite{HNSprd}
\begin{eqnarray}
\sin2\Phi_{B_s} = -0.5\ {\rm to}\ -0.7,
 \ \ \ {\rm (4th\ generation)}
 \label{eq:sin2PhiBs4th}
\end{eqnarray}
where even the sign is predicted. Basically, the range can be
demonstrated by using the (stringent) $\Delta m_{B_s}$ vs (less
stringent) ${\cal B}(B\to X_s\ell^+\ell^-)$ constraints alone,
with $\Delta {\cal A}_{K\pi}$ selecting the minus sign in
Eq.~(\ref{eq:sin2PhiBs4th}), as can be read off from
Fig.~\ref{fig:sin2PhiBs4th}. Note that for different
$m_{t^\prime}$, it maps into a different $\phi_{sb}$-$r_{sb}$
range, with little change in predicted range for
$\sin2\Phi_{B_s}$.

We stress that Eq.~(\ref{eq:sin2PhiBs4th}) can be probed even
before LHCb gets first data, and should help motivate the Tevatron
experiments. {\it It's not over until it's over}, and 2009-2010
could be rather interesting indeed.

\subsubsection{Recent Progress at Tevatron}

We already have a glimpse of what lies ahead in 2008$\,$!

\begin{figure}[t!]
\vskip0.2cm\hskip-0.15cm
\includegraphics[width=0.24\textwidth,height=0.22\textwidth,angle=0]{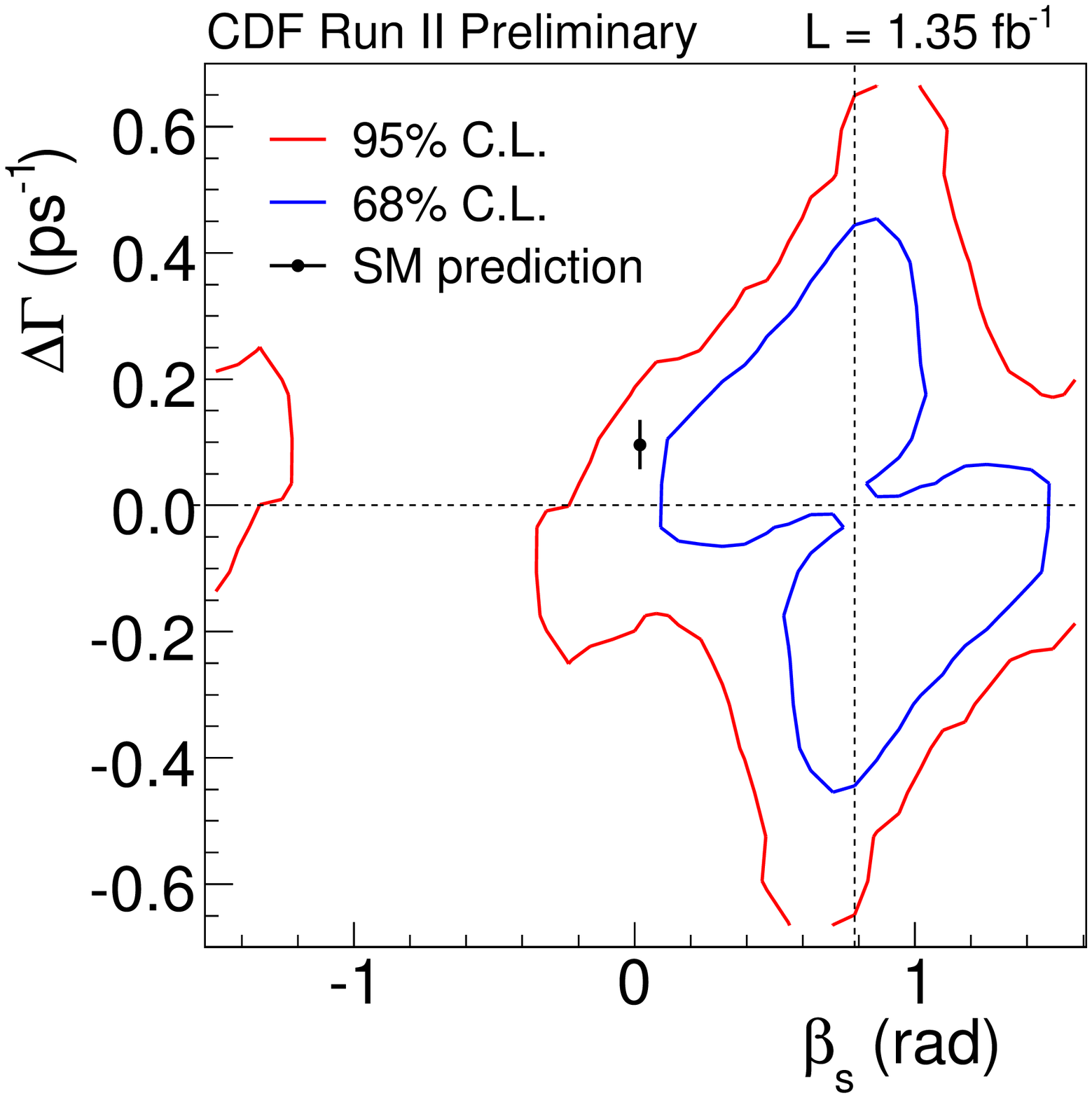}
\vskip-3.52cm\hskip4.1cm
\includegraphics[width=0.25\textwidth,height=0.18\textwidth,angle=0]{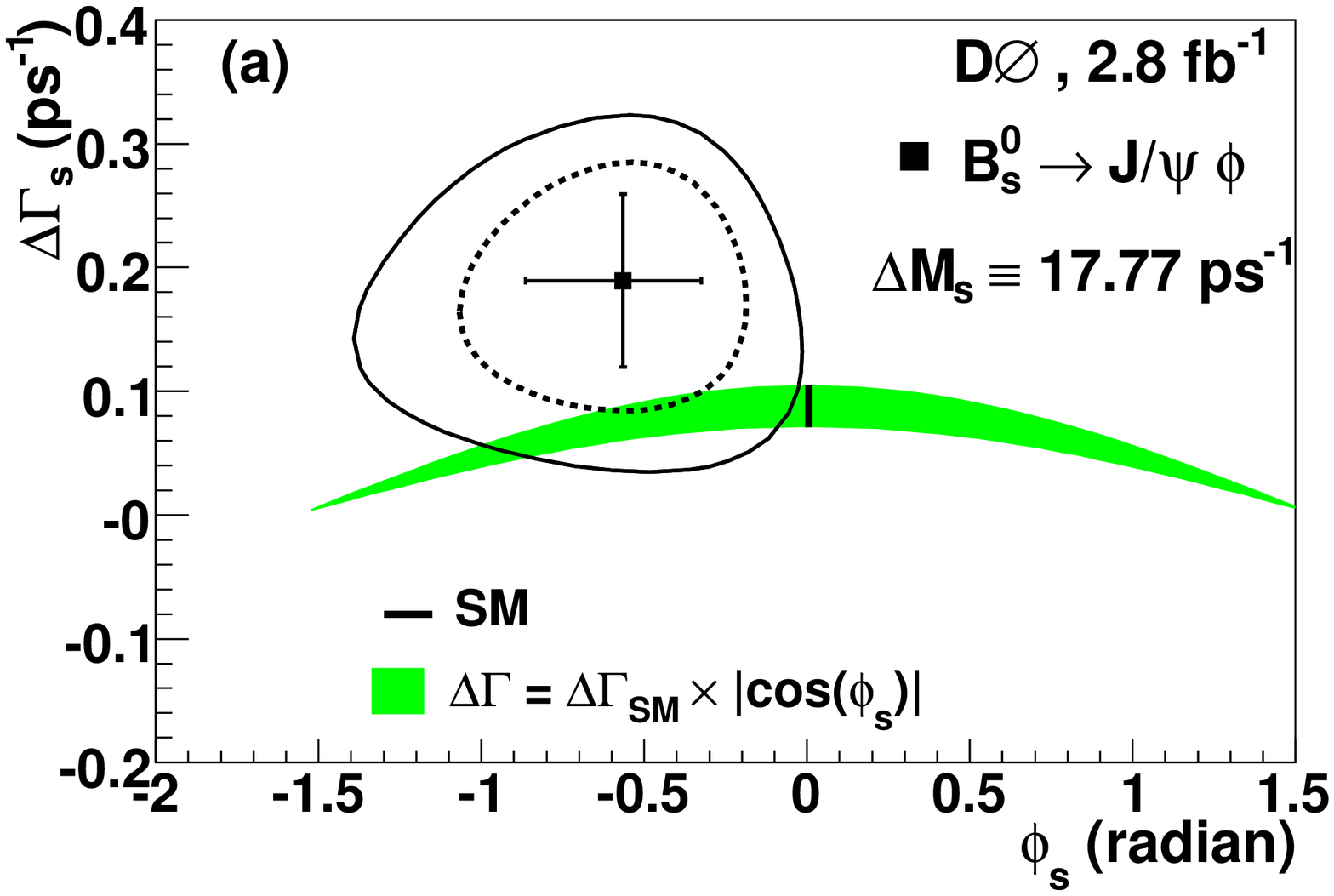}
\vskip0.35cm
 \caption{
 $\Delta\Gamma_{B_s}$ vs $\Phi_{B_s}$ from recent tagged time-dependent studies by
 CDF~\cite{phisCDFtag} using 1.35 fb$^{-1}$ data, and
 D$\emptyset$~\cite{phisDzerotag} using 2.8 fb$^{-1}$ data.
  }
 \label{fig:phis2008W}       
\end{figure}

Using 1.35 fb$^{-1}$ data, CDF has performed the first tagged and
angular-resolved time-dependent CPV study of $B_s \to J/\psi
\phi$. The result~\cite{phisCDFtag}, in terms of $\beta_s =
-\Phi_{B_s}$, is shown in Fig.~\ref{fig:phis2008W}.
A similar analysis has been conducted by D$\emptyset$, assuming
Eq.~(\ref{eq:DeltamBs}) for $\Delta m_{B_s}$ as input. The
result~\cite{phisDzerotag} ($\phi_s = \Phi_{B_s}$), using 2.8
fb$^{-1}$, is also shown in Fig.~\ref{fig:phis2008W}. Up to a
two-fold ambiguity in the CDF result, to the eye, one sees that
both experiments find $\Phi_{B_s}$ to be negative, and is more
consistent with the 4th generation prediction of
Eq.~(\ref{eq:sin2PhiBs4th}), than with the SM prediction of
Eq.~(\ref{eq:sin2PhiBsSM}).

The UTfit group has made the bold attempt to combine the results
of $\Delta m_{B_s}$ as well as Figs.~\ref{fig:phisDzero} and
\ref{fig:phis2008W}, to claim~\cite{phisUTfit} {\it first
evidence} (3.7$\sigma$) for New Physics in $b\leftrightarrow s$
transitions, $\Phi_{B_s} = -19.9^\circ \pm 5.6^\circ$, or
\begin{eqnarray}
\sin2\Phi_{B_s} = -0.64_{-0.14}^{+0.16},
 \ \ \ {\rm (UTfit\ of\ Tevatron\ data)}
 \label{eq:sin2PhiBsW}
\end{eqnarray}
which is tantalizingly consistent with
Eq.~(\ref{eq:sin2PhiBs4th}), the prediction of the 4th generation
model that combined $\Delta m_{B_s}$ and $\Delta {\cal A}_{K\pi}$
results$\,$! The significance is much better than estimated in
Table 1, maybe in part because it contains information beyond $B_s
\to J/\psi \phi$ TCPV analysis. But one should wait for an
official Tevatron average.

Whether measurements with LHC data become available or not, much
progress is expected in the coming year or two, so we will leave
things as it is. We note that models like squark-gluino loops, or
$Z^\prime$ models with specially chosen couplings, could also give
large $\sin2\Phi_{B_s}$, but they would be unable to link with
$\Delta {\cal A}_{K\pi}$.

\subsection{\boldmath ${\cal A}_{\rm CP}(B^+\to J/\psi K^+)$}
 \label{sec:ApsiK+}

Suppose there is New Physics in the $B^+\to K^+\pi^0$ electroweak
penguin. Rather than turning into a $\pi^0$, the $Z^*$ from the
effective $bsZ^*$ vertex could turn into a $J/\psi$ as well. One
can then contemplate DCPV in $B^+\to J/\psi K^+$ as a probe of NP.

$B^+\to J/\psi K^+$ decay is of course dominated by the
colour-suppressed $b\to c\bar cs$ amplitude, which is proportional
to the CKM element product $V_{cs}^*V_{cb}$ that is real to very
good approximation. At the loop level, the penguin amplitudes are
proportional to $V_{ts}^*V_{tb}$ in the SM. Because
$V_{us}^*V_{ub}$ is very suppressed, $V_{ts}^*V_{tb} \cong -
V_{cs}^*V_{cb}$ is not only practically real
(Eq.~(\ref{eq:VtsVtbSM})), it has the same phase as the tree
amplitude. Hence, it is commonly argued that DCPV is less than
$10^{-3}$ in this mode, and $B^+\to J/\psi K^+$ has often been
viewed as a calibration mode in search for DCPV. However, because
of possible hadronic effects, there is no firm prediction that can
stand scrutiny.
A recent calculation~\cite{LiMish07} of $B^0\to J/\psi K_S$ that
combines QCDF-improved factorization and the PQCD approach
confirms the 3 generation SM expectation that ${\cal A}_{\rm
CP}(B^+\to J/\psi K^+)$ should be at the $10^{-3}$ level. Thus, if
\% level asymmetry is observed in the next few years, it would
support the scenario of New Physics in $b\to s$ transitions, while
stimulating theoretical efforts to compute the strong phase
difference between $C$ and $P_{\rm EW}$.

We shall argue that, in the 4th generation scenario, DCPV in
$B^+\to J/\psi K^+$ decay could be at the $\%$ level.

Experiment so far is consistent with zero, but has a somewhat
checkered history. Belle has not updated from their 2003 study
based on 32M $B\bar B$ pairs, although they now have more than
20$\times$ the data. BaBar's study flipped sign from the 2004
study based on 89M, to the 2005 study based on 124M, which seemed
dubious at best. However, the sign was flipped back in PDG 2007,
simply because it was found that the 2005 paper used the opposite
convention to the (standard) one used for 2004. The opposite sign
between Belle and BaBar suppresses the central value, but the
error is at 2\% level. This rules out, for example, the
suggestion~\cite{WS00} of enhanced $H^+$ effect at 10\% level.

One impediment to higher statistics B factory studies is the
systematic error, and it seems difficult to break the 1\% barrier.
Recent progress has been made, however, by D$\emptyset$. Based on
2.8 fb$^{-1}$ data, D$\emptyset$ measures~\cite{ApsiK+_Dzero}
\begin{eqnarray}
{\cal A}_{B^+\to J/\psi K^+}
 = 0.75 \pm 0.61 \pm 0.27\ \%. \ \ \ ({\rm D}\emptyset)
 \label{eq:ApsiK+}
\end{eqnarray}
We should note that there is a correction twice as large as the
value in Eq.~(\ref{eq:ApsiK+}) for the $K^\pm$ asymmetry due to
detector effects, because of its matter composition. Despite this,
of special note is the rather small (roughly a quarter \%$\,$!)
systematic error. This is because one enjoys a larger control
sample in hadronic production, as compared with B factories, e.g.
in $D^*$ tagged $D^0\to K^-\pi^+$ decays. Thus, even scaling up to
8 fb$^{-1}$, one is still statistics limited, and 2$\sigma$
sensitivity for \% level asymmetries could be attainable. CDF
should have similar sensitivity, and the situation can drastically
improve with LHCb data once it becomes available.

\begin{figure}[t!]
\hskip0.9cm
\includegraphics[width=0.30\textwidth,height=0.17\textwidth,angle=0]{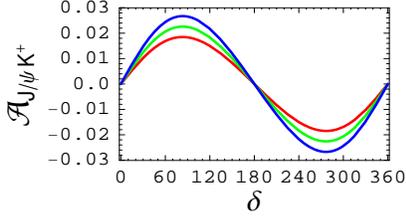}
\vskip0.35cm
 \caption{
 ${\cal A}_{B^+\to J/\psi K^+}$ vs strong phase difference $\delta$ between
 $C$ and $P_{\rm EW}$ in the 4th generation model.
 A nominal $\delta \sim 30^\circ$ is expected from strong phases in
 $J/\psi K^*$ mode. Negative asymmetries are ruled out by the D$\emptyset$
 result of Eq.~(\ref{eq:ApsiK+})
 }
 \label{fig:ApsiK+}       
\end{figure}

The Tevatron study was in fact inspired by a 4th generation
study~\cite{HNSpsiK+}, following the lines that have already been
presented in the previous sections. The 4th generation parameters
are taken from the $\Delta {\cal A}_{K\pi}$ study~\cite{HNS05}. By
analogy with what is observed in $B \to D\pi$ modes, and
especially between different helicity components in $B\to J/\psi
K^*$ decay, the dominant colour-suppressed amplitude $C$ for
$B^+\to J/\psi K^+$ would likely possess a strong phase of order
$30^\circ$. The $P_{\rm EW}$ amplitude is assumed to factorize and
hence does not pick up a strong phase. Heuristically this is
because the $Z^*$ produces a small, colour singlet $c\bar c$ that
penetrates and leaves the hadronic ``muck" without much
interaction, subsequently projecting into a $J/\psi$ meson. With a
strong phase in $C$ and a weak phase in $P_{\rm EW}$, one then
finds ${\cal A}_{B^+\to J/\psi K^+} \simeq \pm 1\%$.

We plot ${\cal A}_{B^+\to J/\psi K^+}$ vs phase difference
$\delta$ in Fig.~\ref{fig:ApsiK+}, with $\phi_{sb}$ fixed to the
range corresponding to Eq.~(\ref{eq:sin2PhiBs4th}), and notation
as in Fig.~\ref{fig:sin2PhiBs4th}. Negative sign is ruled out by
Eq.~(\ref{eq:ApsiK+}).
But of course, DCPV is directly proportional to the strong phase
difference, which is not well predicted.

We remark that models like $Z^\prime$ with FCNC couplings could
also generate various effects we have discussed. For example, with
$\delta \sim 30^\circ$, ${\cal A}_{B^+\to J/\psi K^+}$ could be
considerably larger than a percent. With the D$\emptyset$ result
of Eq.~(\ref{eq:ApsiK+}), however, only \% level asymmetries are
allowed, ruling out a large (and in any case quite arbitrary)
region of parameter space for $Z^\prime$ effects.

\section{\boldmath $H^+$ Probes}
\label{sec:H+}

When $b\to s\gamma$ was first announced by CLEO~\cite{bsgammaCLEO}
in 1994 with 3 fb$^{-1}$ data on the $\Upsilon(4S)$, it provided
one of the most powerful constraints on many kinds of New Physics
that enter the loop. Here we illustrate the stringent bound it
provides on the charged Higgs boson $H^+$ that automatically
exists in minimal SUSY. A second probe of $H^+$, becoming relevant
only recently at the B factories is, surprisingly, a tree level
effect in $B^+\to \tau^+\nu$.

\subsection{\boldmath $b\to s\gamma$}
\label{sec:btosgamma}

The inclusive $b\to s\gamma$ decay, identified with $B\to
X_s\gamma$ experimentally, where $X_s$ are
reconstructed~\cite{bsgammaCLEO} as $K+n\pi$ (partial
reconstruction), is one of the most important probes of NP. There
had been good agreement for the past few years between NLO theory
and the experimental average of~\cite{HFAG}
\begin{eqnarray}
{\cal B}_{B\to X_s\gamma} = (3.55\pm0.26) \times 10^{-4}\
 {\rm (HFAG\;06)},
 \label{eq:Xsgamma}
\end{eqnarray}
which has gone beyond partial reconstruction, but on background
reduction after selecting an energetic photon.

Recently, however, the NNLO theory prediction has shifted
lower~\cite{Misiak07,BN07} to $\sim 3 \times 10^{-4}$, with errors
comparable to experiment. Although the NNLO work continues, the
ball appears to be in the experiments' court.

\begin{figure}[t!]
\hskip1.75cm
\includegraphics[width=0.28\textwidth,height=0.26\textwidth,angle=0]{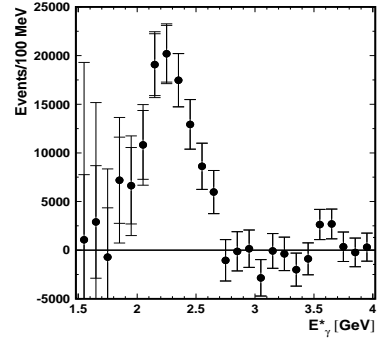}
 \vskip0.5cm
\caption{
 The $E_\gamma$ spectrum above 1.8 GeV in $\Upsilon(4S)$ frame
 for inclusive $b\to s\gamma$
 (Belle~\cite{bsgamma_Belle04} 152M $B\bar B$ pairs).}
 \label{fig:Egamma}       
\end{figure}

To improve on the experimental error, besides an ever larger
dataset, the photon energy cut, e.g. $E_\gamma > 1.8$ GeV (see
Fig.~\ref{fig:Egamma}) in the Belle study
using~\cite{bsgamma_Belle04} 152M $B\bar B$ pairs, should be
lowered further. Dutifully, Belle has just come
out~\cite{bsgamma_Belle08} with a new analysis using 657M $B\bar
B$ pairs, while managing to lower $E_\gamma$ cut to 1.7 GeV.
Agreement with theory is slightly improved.

To confront the theoretical advancement, however, a fresher
approach may eventually be needed.
A promising new development, as the B factories increase in data,
is the full reconstruction of the tag side $B$ meson. The signal
side is then just an energetic photon, without specifying the
$X_s$ system. The systematics would be quite different from
previous approaches. A first attempt has been performed by
BaBar~\cite{bsgamma_BaBar08} recently. But since full
reconstruction takes a $10^{-3}$ hit in efficiency, it seems that
the NNLO theory development would demand a Super B factory upgrade
to continue the supreme dialogue between theory vs experiment in
this mode.

This close dialogue allowed $b\to s\gamma$ to provide one of the
most stringent bounds on NP models. The process is sensitive to
all types of possible NP in the loop, such as stop-charginos,
where a large literature exists. However, $b\to s\gamma$ is best
known for its stringent constraint on the MSSM (minimal SUSY SM)
type of $H^+$ boson. Furthermore, the SUSY related studies all
need mechanisms to cancel against the large charged Higgs effect.
We therefore focus on the $H^+$ effect in the loop.

MSSM demands at least two Higgs doublets (2HDM), where one Higgs
couples to right-handed down type quarks, the other to up type.
The physical $H^+$ is a cousin of the $\phi_{W^+}$ Goldstone boson
of the SM
that gets eaten by the $W^+$. It is the $\phi_{W^+}$ that couples
to masses, and is at the root of the nondecoupling phenomenon of
the heavy top quark in the loop. In $bs\gamma$ coupling, however,
the top is effectively decoupled (less than logarithmic dependence
on $m_t$), by a subtlety of gauge invariance. This underlies the
reason why QCD corrections make such large
impact~\cite{bsgammaQCD} in this loop-induced decay. It also makes
the process sensitive to NP such as $H^+$.

\begin{figure}[t!]
\vskip0.1cm\hskip1.2cm
\includegraphics[width=0.35\textwidth,height=0.22\textwidth,angle=0]{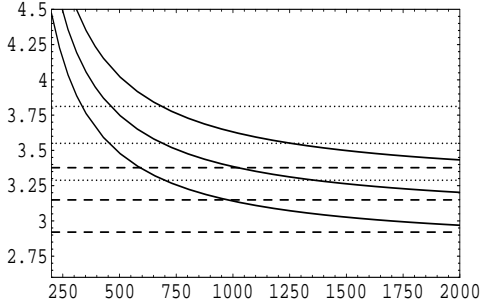}
 \vskip0.4cm
\caption{
 ${\cal B}(B\to X_s\gamma)$ vs $m_{H^+}$ in MSSM type two Higgs
 doublet model, with $\tan\beta = 2$ (taken from~\cite{Misiak07}).
 For large $m_{H^+}$,
 one approaches SM (dashed lines), while for low $m_{H^+}$ there
 is great enhancement. Dotted lines is experimental range.
 }
 \label{fig:Misiak}       
\end{figure}

Replacing the $W^+$ by $H^+$ in the loop, in the MSSM type of
2HDM, the $H^+$ effect {\it always enhances $b\to s\gamma$ rate,
regardless of $\tan\beta$}, which was pointed out 20 years
ago~\cite{GW88,HW88}, where $\tan\beta$ is the ratio of v.e.v.s
between the two doublets. Basically, the $H^+$ couples to
$m_t\cot\beta $ at one end of the loop, and to $-m_b\tan\beta$ on
the other end, so this contribution is independent of $\tan\beta$,
and the sign is fixed to be always constructive with the SM
amplitude\footnote{
 In the other type of 2HDM, where both $u$ and $d$ quarks
 gets mass from the same Higgs doublet,
 the $H^+$ effect is destructive~\cite{HW88}.
 }.

We take the plot from Ref.~\cite{Misiak07}, where the  NNLO result
of ${\cal B}(B\to X_s\gamma)$ vs $m_{H^+}$ is compared with
data~\cite{HFAG}. A nominal $\tan\beta = 2$ is taken. By comparing
the lower range of NNLO result with the higher range of
Eq.~(\ref{eq:Xsgamma}), one has the bound
\begin{eqnarray}
m_{H^+} > 295\ {\rm GeV\ \ \ ({\rm NNLO+HFAG06})},
 \label{eq:H+bound}
\end{eqnarray}
at 90\% C.L. If one takes the central value of both results
seriously, one could say~\cite{Misiak07} that an $H^+$ boson with
mass around 695 GeV is needed to bring the NNLO rate up to
Eq.~(\ref{eq:Xsgamma}). Again, this is because the $H^+$ effect in
the MSSM type of 2HDM is always constructive~\cite{HW88} with
$\phi_W$ effect in SM.
%

The ongoing saga should be watched. It would be interesting with
LHC turn on, especially if a charged Higgs boson is discovered.
Much more information could be extracted in the future with a
Super B Factory.

\subsection{\boldmath $B\to \tau\nu$ and $D^{(*)}\tau\nu$ }
\label{sec:btotaunu}

\subsubsection{$B\to \tau\nu$ Meausurement}

As a cousin of the $\phi_{W^+}$, the $H^+$ boson has an amazing
tree level effect that has only recently come to fore by the
prowess of the B factories.

Like $\pi^+,\ K^+ \to \ell^+\nu_\ell$ decay, one has the formula
for $B^+\to \tau^+\nu_\tau$ decay,
\begin{eqnarray}
 {\cal B}_{B\to \tau\nu} 
 = r_H \frac{G_F^2 m_B m_\tau^2}{8\pi}
       \left[1-\frac{m_\tau^2}{m_B^2}\right]
       \tau_{B} f_B^2\vert V_{ub}\vert^2,
 \label{eq:Btaunu}
\end{eqnarray}
where $r_H = 1$ for SM, but~\cite{Hou93}
\begin{eqnarray}
 r_H = \left[1-\frac{m_{B^+}^2}{m_{H^+}^2}\tan^2\beta\right]^2,
 \label{eq:rH}
\end{eqnarray}
for 2HDM.
Within SM, the pure gauge $W^+$ effect is helicity suppressed,
hence the effect vanishes with the $m_\tau$ mass. For $H^+$, there
is no helicity suppression, but one has the ``Higgs affinity"
factor, i.e. mass dependent couplings. With $m_u$ negligible, the
$H^+$ couples as $m_\tau m_b \tan^2\beta$. This leads to the $r_H$
factor of Eq.~(\ref{eq:rH}), where the sign between the SM and
$H^+$ contribution is always destructive~\cite{Hou93}.

\begin{figure}[t!]
\hskip1.9cm
\includegraphics[width=0.30\textwidth,height=0.21\textwidth,angle=0]{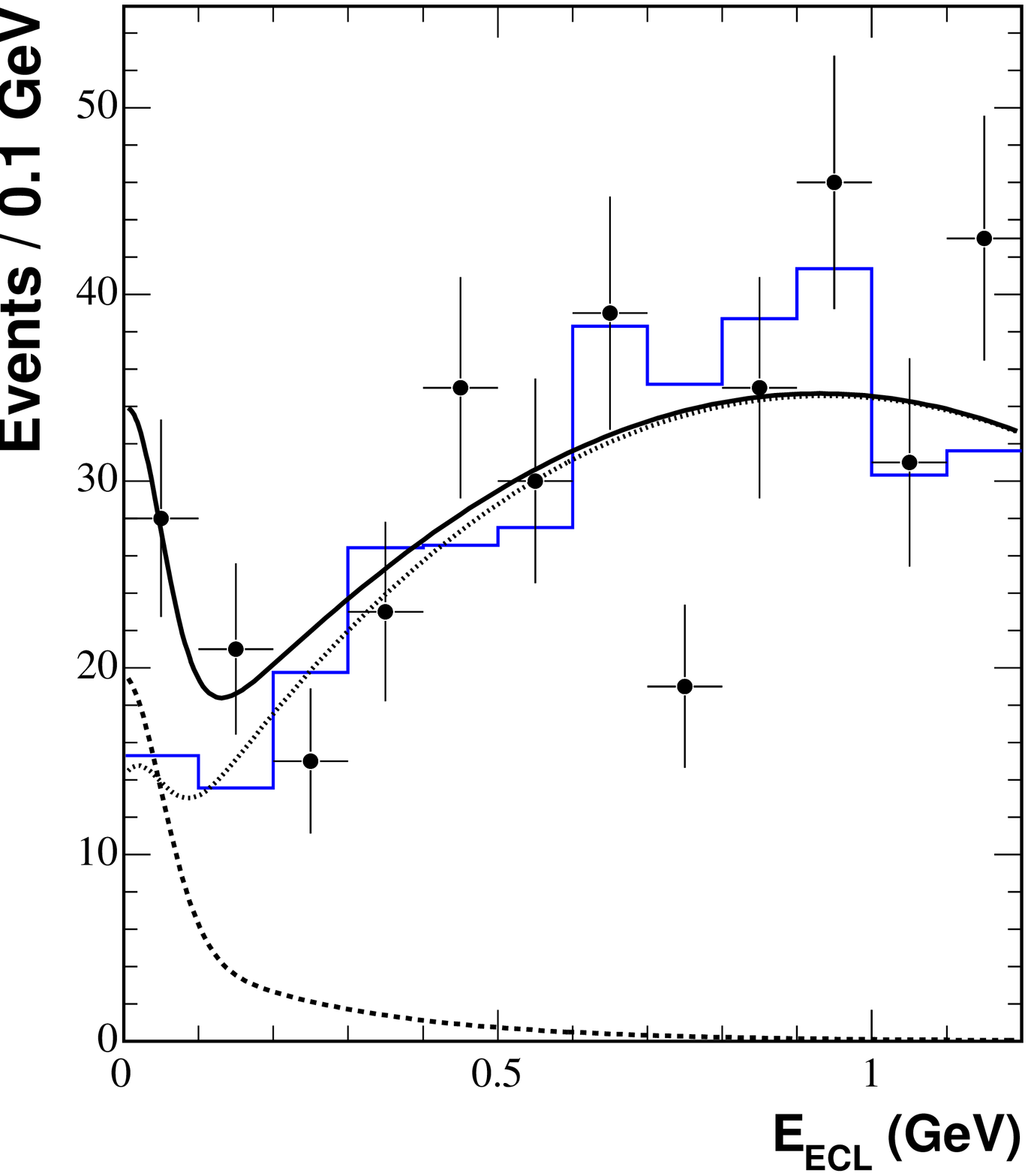}
\vskip-0.07cm\hskip1.57cm
\includegraphics[width=0.315\textwidth,height=0.22\textwidth,angle=0]{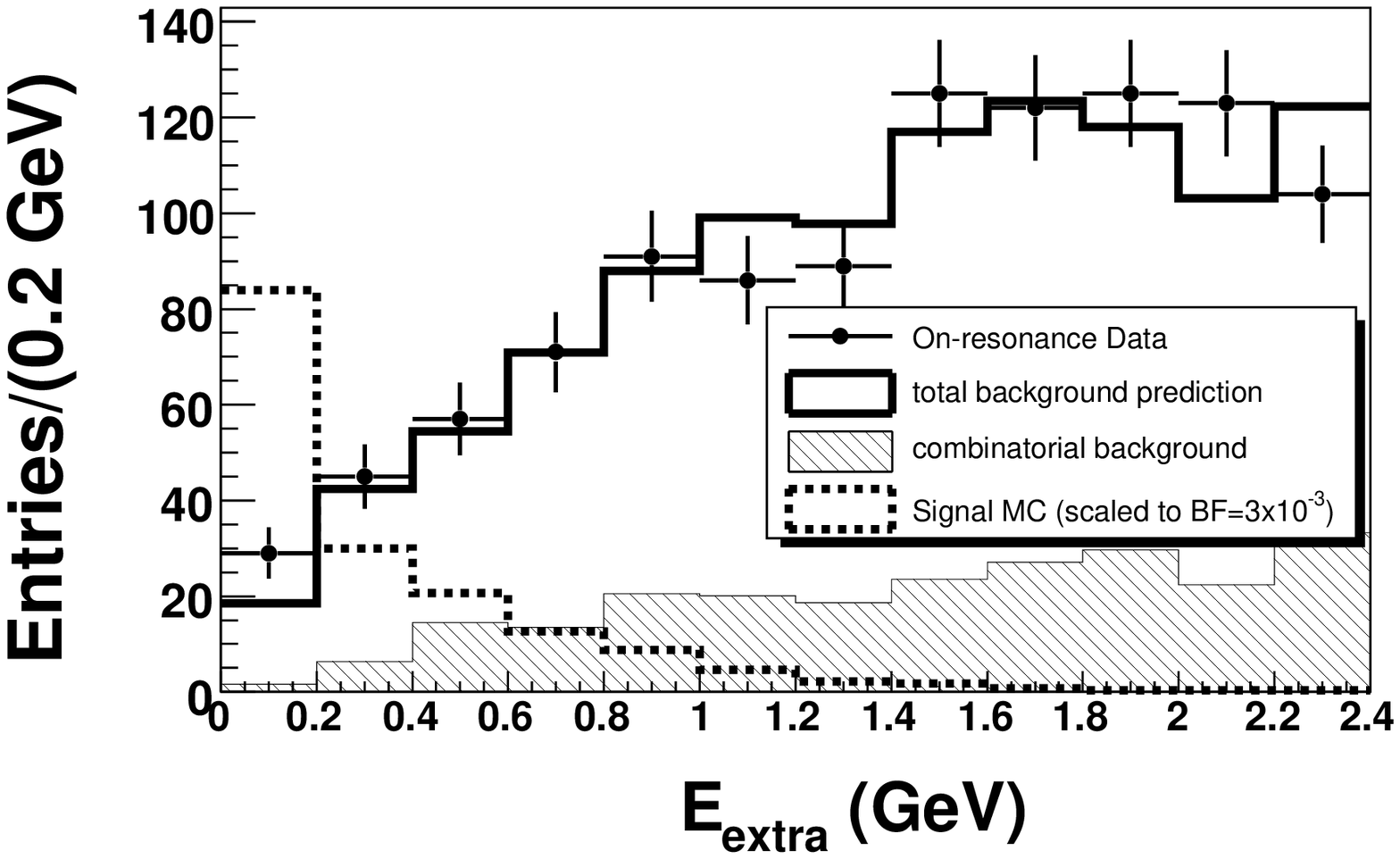}
\vskip0.2cm \caption{
 Data showing evidence for $B\to\tau\nu$ (hadronic tag) search
 by Belle~\cite{taunu_Belle06} and BaBar~\cite{taunu_BaBar07}.}
 \label{fig:taunu}       
\end{figure}

$B^+\to \tau^+\nu$ followed by $\tau^+$ decay results in at least
two neutrinos, which makes background very hard to suppress in the
$B\bar B$ production environment. Thus, for a long time, the limit
on $B^+\to \tau^+\nu$ was rather poor and not so interesting. This
had allowed for the possibility that the effect of the $H^+$ could
even dominate over SM,\footnote{
 In fact, the $H^+$ effect had originally been used to
 enhance~\cite{GH1} $b\to c\tau\nu$ rate to 10\% level,
 in an attempt to account for a discrepancy in
 the measured semileptonic branching ratio.
 This possibility was subsequently ruled out by experimental
 measurement~\cite{PDG} of $b\to c\tau\nu$ to be at SM expectation level.
 }
given that the SM expectation was only at $10^{-4}$ level. The
change came with the enormous number of B mesons accumulated by
the B factories, allowing the aforementioned full reconstruction
method to become useful.

Fully reconstructing the tag side B meson in, e.g. $B^-\to
D^0\pi^-$ decay, one has an efficiency of only 0.1\%--0.3\%. At
this cost, however, one effectively has a ``B beam".
As shown in Fig.~\ref{fig:taunu}, using full reconstruction in
hadronic modes and with a data consisting of 449M $B\bar B$ pairs,
in 2006 Belle found $17.2^{+5.3}_{-4.7}$ events, where the $\tau$
decay was searched for in decays to $e\nu\nu$, $\mu\nu\nu$,
$\pi\nu$ and $\rho\nu$ modes. This constituted the first evidence
(at 3.5$\sigma$) for $B^+\to \tau^+\nu$, with~\cite{taunu_Belle06}
\begin{eqnarray}
  {\cal B}_{B\to \tau\nu}  = (1.79^{+0.56+0.46}_{-0.49-0.51}) \times 10^{-4}\;\
 {\rm (Belle\;449M)}.
 \label{eq:taunuBelle}
\end{eqnarray}
With 320M $B\bar B$s and $D\ell\nu$ reconstruction on tag side,
however, BaBar saw no clear signal, giving
$(0.88^{+0.68}_{-0.67}\pm0.11) \times 10^{-4}$.
Updating more recently to 383M, the $D\ell\nu$ tag result of
$(0.9\pm0.6\pm0.1) \times 10^{-4}$ is not different from the 320M
result. However, with hadronic tag, BaBar now also reports some
evidence, at $(1.8^{+0.9}_{-0.8}\pm0.4\pm0.2) \times 10^{-4}$
(Fig.~\ref{fig:taunu}), which is quite consistent with the Belle
result of Eq.~(\ref{eq:taunuBelle}). The combined result for BaBar
is~\cite{taunu_BaBar07},
\begin{eqnarray}
  {\cal B}_{B\to \tau\nu}
   = (1.2\pm0.4\pm0.36) \times
   10^{-4}\ 
 {\rm (BaBar\;383M)},
 \label{eq:taunuBaBar}
\end{eqnarray}
where we have followed HFAG to combine the background and
efficiency related errors. Eq.~(\ref{eq:taunuBaBar}) has
2.6$\sigma$ significance, and is diluted by the semileptonic tag
measurement, but it is basically consistent with the Belle result.

%

Taking central values from lattice for $f_B$, and $\vert
V_{ub}\vert$ from semileptonic decays, the nominal SM expectation
is $(1.6\pm0.4) \times 10^{-4}$. Thus, Belle and BaBar have
reached SM sensitivity, and Eqs.~(\ref{eq:taunuBelle}) and
(\ref{eq:taunuBaBar}) now place a constraint on the
$\tan\beta$-$m_{H^+}$ plane through $r_H \simeq 1$.
If one has a Super B factory, together with development of lattice
QCD, this can become a superb probe of the $H^+$, complementary to
direct $H^+$ searches at the LHC.
%
A particularly nice feature is its theoretical cleanliness, all
hadronic effects being summarized in $f_B$.

\subsubsection{$B\to D^{(*)}\tau\nu$ Meausurement}

An analogous mode with larger branching ratio, $B\to
D^{(*)}\tau\nu$, has recently emerged. In 2007 Belle announced the
observation of~\cite{Dstartaunu_Belle07}
\begin{eqnarray}
  {\cal B}_{D^{*-}\tau\nu} = (2.02^{+0.40}_{-0.37}\pm0.37)\; \%\;\
 {\rm (Belle\;535M)},
 \label{eq:DstartaunuBelle}
\end{eqnarray}
based on $60^{+12}_{-11}$ reconstructed signal events, which is a
5.2$\sigma$ effect. Subsequently, based on 232M $B\bar B$ pairs,
BaBar announced the observation (over 6$\sigma$) of
$D^{*0}\tau\nu$, and evidence (over 3$\sigma$) for
$D^{+}\tau\nu$~\cite{Dstartaunu_BaBar07}
\begin{eqnarray}
  {\cal B}_{D^{*0}\tau\nu} &=& (1.81\pm0.33\pm0.11\pm0.06)\; \% \nonumber \\
  {\cal B}_{D^+\tau\nu} &=& (0.90\pm0.26\pm0.11\pm0.06)\; \%\;\
 \label{eq:DstartaunuBaBar} \\
 &&{\rm \hskip2cm (BaBar\;232M)}, \nonumber
 \label{eq:DstartaunuBaBar}
\end{eqnarray}
where the last error is from normalization.

The SM branching ratios, at 1.4\% for $B\to D^{*}\tau\nu$, are
poorly estimated. Furthermore, though the $H^+$ could hardly
affect the $B \to D^{*}\tau\nu$ rate, it could leave its mark on
the $D^*$ polarization. The $B \to D\tau\nu$ rate, like $B\to
\tau\nu$ itself, is more directly sensitive to $H^+$~\cite{GH2}.
More theoretical work, as well as polarization information, would
be needed for BSM (in particular, $H^+$ effect) interpretation.
But it is rather curious that, almost 25 years after the first B
meson was reconstructed, we have a newly measured mode with $\sim$
2\% branching faction!

\subsubsection{Comment on New Physics in $D_s^+\to \mu^+\nu,\ \tau^+\nu$}

The process $D_s^+\to \ell^+\nu$, where $\ell = \mu,\ \tau$,
proceeds via $c\bar s$ annihilation, and the decay branching ratio
formula is very similar to Eq.~(\ref{eq:Btaunu}), with $r_H$ set
to 1 in SM. Since this is a tree level process proceeding without
CKM suppression, New Physics effect through the charged Higgs is
expected to be small~\cite{Hou93}. The rate measures $f_{D_s}\vert
V_{cs}\vert$ in a rather clean way.

The experimental measurement has become rather
precise~\cite{fDsCLEO07,fDsBelle08} recently
\begin{eqnarray}
  f_{D_s}\vert^{\rm expt} &=& 277 \pm 9\ {\rm MeV},
 \label{eq:fDsExp}
\end{eqnarray}
assuming $\vert V_{cs}\vert = 1$. Given confirmation between two
experiments,\footnote{
 We note that the BaBar measurement~\cite{fDsBaBar07} is not an
 absolute branching ratio measurement. But the result is similar
 in any case.}
there is little likelihood that the experimental number would
change much.

The experimental result has been compared~\cite{Kronfeld} recently
with a {\it very precise} result from the lattice~\cite{latt08},
\begin{eqnarray}
  f_{D_s}\vert^{\rm latt} &=& 241 \pm 3\ {\rm MeV},\ \
   ({\rm ``rooting"})
 \label{eq:fDsLatt}
\end{eqnarray}
Note the \% level errors! This precision arises in the staggered
fermion approach in lattice QCD, with a big assumption to simplify
the computation of the fermion determinant, called ``rooting".
Ref.~\cite{Kronfeld} claims that the precision of
Eq.~(\ref{eq:fDsLatt}) can stand scrutiny, then goes on to claim
that this discrepancy suggests New Physics.

It is not our purpose to go into detail or comment on the
intricacies of lattice QCD computations, although we have used the
discrepancy of the above two equations to argue, in an intuitive
way, that $B_s$ mixing in SM is likely to be larger than the
experimental measurement of Eq.~(\ref{eq:DeltamBs}). But we do
find the claim of Ref.~\cite{Kronfeld} incredulous. The percent
level numerical accuracy of a lattice calculation should be
scrutinized thoroughly by the lattice QCD community before such a
claim can be made. Afterall, unlike the experimental situation,
the lattice result of Eq.~(\ref{eq:fDsLatt}) is so far a
stand-alone result. Furthermore, the New Physics ``models"
proposed by Ref.~\cite{Kronfeld} are rather {\it ad hoc} and
constructed, and not the ones that this brief Review would like to
contemplate.

To paraphrase Einstein, God may not be subtle at all, but
malicious, if the tree dominant and Cabibbo allowed $D_s^+ \to
\ell^+ \nu$ was chosen as the first place to reveal to us signs of
New Physics.

\section{\boldmath Electroweak Penguin: $bsZ$ Vertex, $Z^\prime$, DM}
 \label{sec:EWP}

In Sec.~\ref{sec:DeltaAKpi}, we discussed the effects of the $b\to
s\bar qq$ electroweak penguin interfering with the strong penguin
and tree amplitudes. The quintessential electroweak penguin would
be $b\to s\ell^+\ell^-$ decay, or $b\to s\nu\nu$ that has no
photonic contribution. We now discuss how the study of these
processes, present already in SM, could help us probe New Physics
as well.

\subsection{\boldmath $A_{\rm FB}(B\to K^*\ell^+\ell^-)$}
\label{sec:AFB}

The $B\to K^*\ell^+\ell^-$ process ($b\to s\ell^+\ell^-$ at
inclusive level) arises from photonic penguin, $Z$ penguin and box
diagrams. The top quark exhibits nondecoupling in the latter
diagrams, analogous to the electroweak penguin effect in $B^+\to
K^+\pi^0$, and the box diagrams for $B_s^0$-$\bar B_s^0$ mixing.
It turns out that, due to this nondecoupling effect of the top
quark, the $Z$ penguin dominates the $b\to s\ell^+\ell^-$ decay
amplitude~\cite{HWS87}. Interference between the vector ($\gamma$
and $Z$) and axial vector ($Z$ only) contributions to
$\ell^+\ell^-$ production gives rise to an interesting
forward-backward asymmetry~\cite{AFB91}. This is akin to the
familiar ${\cal A}_{\rm FB}$ in $e^+e^-\to f\bar f$, except the
enhancement of $bsZ$ penguin with respect to $bs\gamma$, brings
the $Z$ much closer to the $\gamma$ in $B$ decay, and one probes
potential New Physics in the loops.

Both the inclusive $B\to X_s\ell^+\ell^-$ and exclusive $B\to
K^{(*)}\ell^+\ell^-$ decays have now been measured~\cite{PDG}.
Interest has turned to ${\cal A}_{\rm FB}$ for $B\to
K^{*}\ell^+\ell^-$. The study for inclusive ${\cal A}_{\rm FB}$ is
more challenging, and largely impossible in hadronic environment.
A commonly used formula for the differential ${\cal A}_{\rm FB}$
is
\begin{eqnarray}
  d\frac{{\cal A}_{\rm FB}(q^2)}{dq^2}
   &\propto& C_{10} \xi(q^2)\left[{\rm Re}(C_9^{\rm eff})F_1 +
  \frac{1}{q^2}C_7^{\rm eff} F_2\right],
 \label{eq:AFB}
\end{eqnarray}
where $C_{i}$ are Wilson coefficients, and formulas for $\xi(q^2)$
and the form factor related functions $F_1$ and $F_2$ can be found
in Ref.~\cite{AFB00}. From the $1/q^2$, it is clear that $C_7$ is
effectively the photon contribution, while $C_9^{\rm eff}$ and
$C_{10}$ are from $Z$ penguin and box diagram. Within SM, these
Wilson coefficients are practically real, as is apparent from the
formula. $C_{9}^{\rm eff}$ receives some long distance $c\bar c$
effect.

\begin{figure}[t!]
\vskip-0.07cm \hskip1.7cm
\includegraphics[width=0.30\textwidth,height=0.21\textwidth,angle=0]{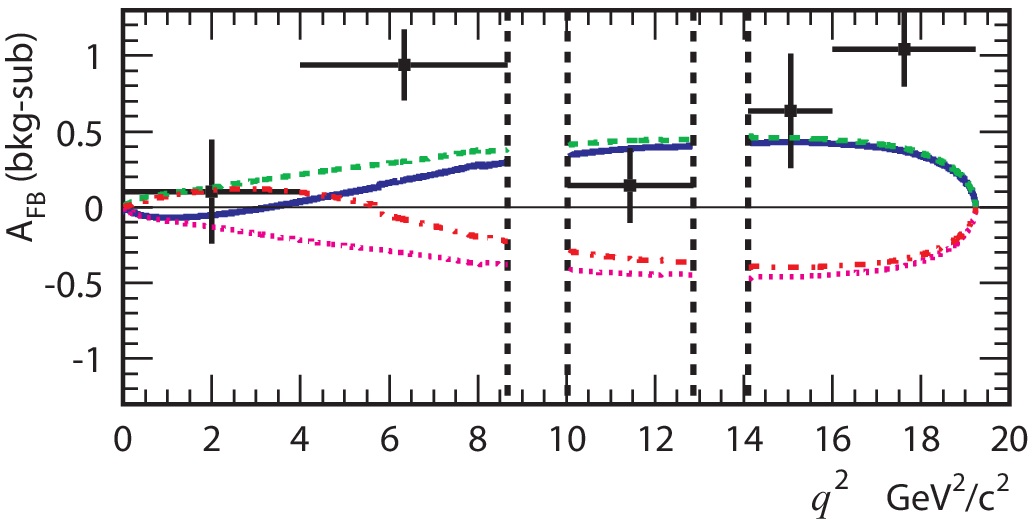}
\vskip0.2cm \hskip1.78cm
\includegraphics[width=0.302\textwidth,height=0.178\textwidth,angle=0]{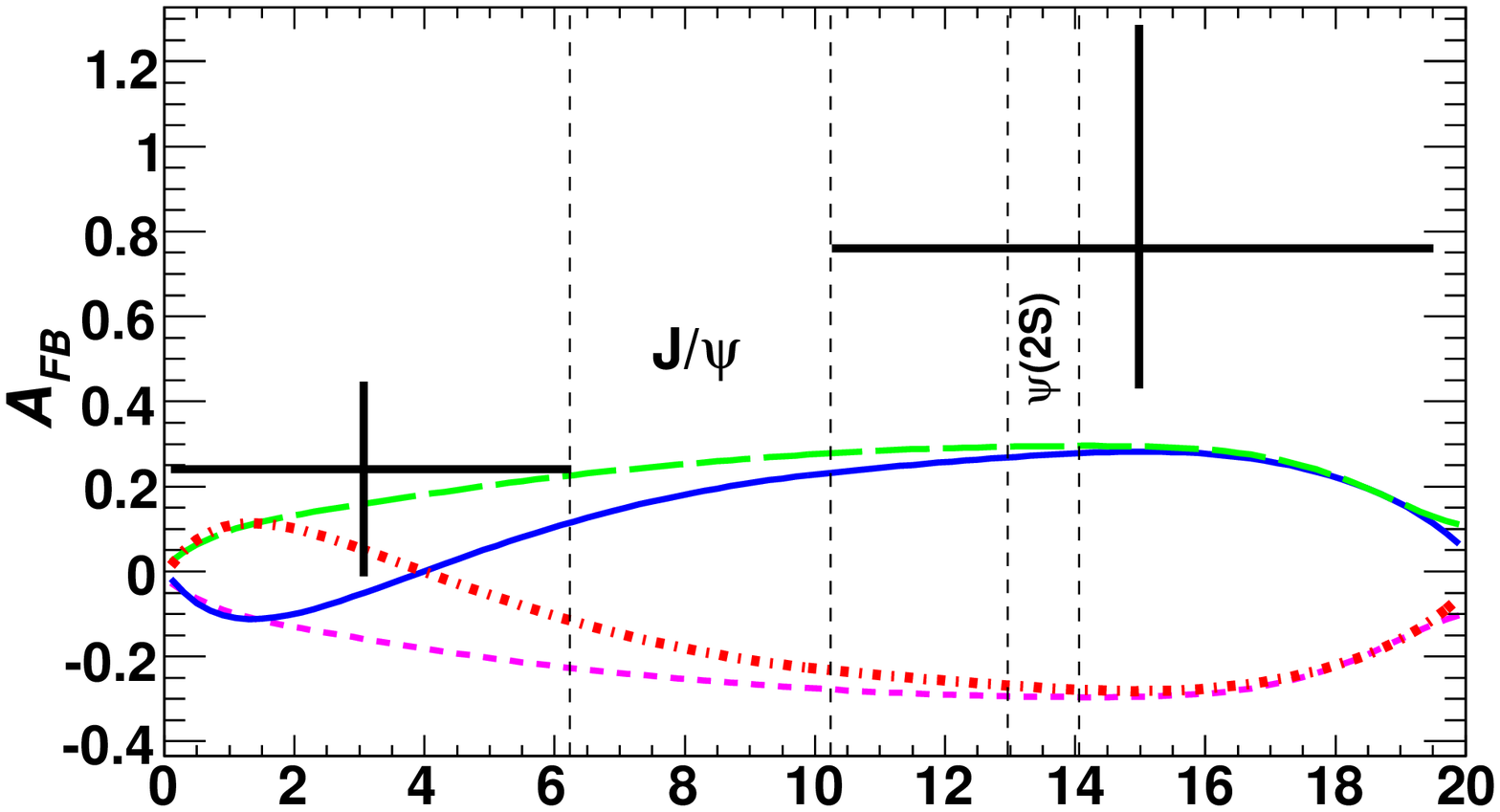}
\vskip0.2cm \caption{
 Measurements of forward-backward asymmetry ${\cal A}_{\rm FB}$
 in $B\to K^{*}\ell^+\ell^-$ by Belle~\cite{AFB_Belle06} and
 BaBar~\cite{AFB_BaBar08}.
 The two lower curves are for flipping the sign of either $C_9$
 or $C_{10}$ with respect to SM (solid curve), while
 the upper curve is for $C_9$ and $C_{10}$ both flipping sign.
 }
 \label{fig:AFBexp}       
\end{figure}

As shown in Fig.~\ref{fig:AFBexp}, the study of forward-backward
asymmetry in $B\to K^*\ell^+\ell^-$ by Belle with 386M $B\bar B$
pairs~\cite{AFB_Belle06} is consistent with SM, and rules out the
possibility of flipping the sign of $C_9$ or $C_{10}$ separately
from SM value, but having both $C_9$ or $C_{10}$ flipped in sign
(equivalent to flipping sign of $C_7$) is not ruled out. BaBar
took the more conservative approach of giving ${\cal A}_{\rm FB}$
in just two $q^2$ bins, below and above $m_{J/\psi}^2$. With 229M,
the higher $q^2$ bin is consistent~\cite{AFB_BaBar06} with SM and
disfavors BSM scenarios. Interestingly, in the lower $q^2$ bin,
while sign-flipped BSM's are less favored, the measurement is
$\sim 2\sigma$ away from SM.

BaBar has just updated to 384M~\cite{AFB_BaBar08}. For the high
$q^2$ bin, the results are qualitatively the same as before. For
the low $q^2$ bin ($4m_\mu^2$ to 6.25 GeV$^2/c^4$), as can be seen
from the second plot in Fig.~\ref{fig:AFBexp}, BaBar has improved
its measurement to ${\cal A}_{\rm FB}\vert_{{\rm low}\ q^2} =
0.24^{+0.18}_{-0.23}\pm0.05$. This compares with the SM
expectation that ${\cal A}_{\rm FB}\vert_{{\rm low}\ q^2}^{\rm SM}
= -0.03\pm0.01$. Though not excluded, viewed together with the
Belle result, it seems that the low $q^2$ behavior is not quite
SM-like.

While the above is interesting, it is clear that the B factory
statistics is still rather limited, and cannot be much improved
without a Super B factory. But LHCb can do very well in this
regard within a couple of years.

In the context of LHCb prospects, it was recently
noticed~\cite{HHM07} that, in Eq.~(\ref{eq:AFB}), there is no
reason {\it a priori} why the Wilson coefficients should be kept
real when probing BSM physics! Note that ${\rm Re}(C_9)$ in
Eq.~(\ref{eq:AFB}) differs from $C_9$ within SM by just a small
correction arising from long distance $c\bar c$ effects. But if
one keeps an open mind (rather than, for example, taking the
oftentimes tacitly assumed Minimal Flavour Conservation mindset),
Eq.~(\ref{eq:AFB}) should be restored to its proper form,
\begin{eqnarray}
   {\rm Re}\left(C_9^{\rm eff}C_{10}^*\right) {\cal F}_1 
   + \frac{1}{q^2}{\rm Re}\left(C_7^{\rm eff}C_{10}^*\right) {\cal F}_2,
 \label{eq:AFBcomplex}
\end{eqnarray}
where ${\cal F}_i$ are form factor combinations. We are not
concerned with $CP$ conserving long distance effects, but the
possibility that the $C_i$s may pick up BSM $CP$ violating phases.
If present, they could enrich the interference pattern through
Eq.~(\ref{eq:AFBcomplex}), compared with Eq.~(\ref{eq:AFB}), which
has practically {\it assumed} real short distance Wilson
coefficients. After all, the equivalent $C_9$ and $C_{10}$ for
$B^+\to K^+\pi^0$ decay seem to carry {\it large} weak phases, if
$P_{\rm EW}$ is the culprit for the $\Delta{\cal A}_{K\pi}$
problem discussed in Sec.~\ref{sec:DeltaAKpi}.
Let Nature speak through data!

\begin{figure}[t!]
\vskip0.2cm\hskip1.5cm
\includegraphics[width=0.34\textwidth,height=0.23\textwidth,angle=0]{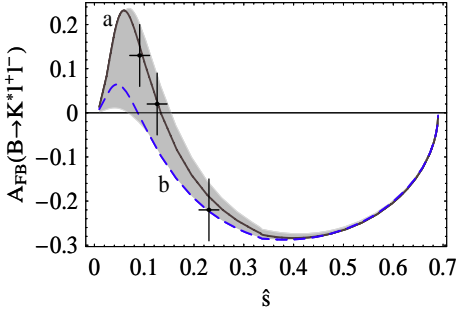}
 \vskip0.2cm
\caption{
 Possible ${\cal A}_{\rm FB}$ in $B\to K^{*}\ell^+\ell^-$ allowed
 by complex Wilson coefficients, Eq.~(\ref{eq:AFBcomplex}). The three data
 points are taken from 2 fb$^{-1}$ LHCb Monte Carlo for illustration,
 which has the power to distinguish between SM (solid curve) vs e.g. fourth
 generation model (dashed curve).}
 \label{fig:AFB_NP}       
\end{figure}

Taking the sign convention of LHCb, which is opposite to Belle and
BaBar, we illustrate~\cite{HHM07} in Fig.~\ref{fig:AFB_NP} the
situation where New Physics enters through effective $bsZ$ and
$bs\gamma$ couplings. In this case, $C_9$ and $C_{10}$ cannot
differ by much at short distance, which is the reason for the
``degenerate tail" for larger $\hat s \equiv q^2/m_B^2$. But by
allowing the Wilson coefficients to be only constrained by the
measured radiative and electroweak penguin rates, ${\cal A}_{\rm
FB}$ could in fact vary in the shaded region, practically for $q^2
< m_{J/\psi}^2$, and not just in the position of the zero. The
fourth generation with parameters as determined from $\Delta
m_{B_s}$, ${\cal B}(B\to X_s\ell^+\ell^-)$ and $\Delta {\cal
A}_{K\pi}$ belongs to this class of BSM models, and is plotted as
the dashed line for illustration. To get a feeling for the future,
we take the MC study~\cite{LHCb2fb} for 2 fb$^{-1}$ data by LHCb
(achievable in a couple years of running) and plot three sample
data points to illustrate expected data quality. These data points
are based on the SM (solid line), and it is clear that LHCb can
distinguish between SM and the 4th generation.

Back to the present. From Fig.~\ref{fig:AFB_NP} we could also
compare with Belle and BaBar data~\cite{AFB_Belle06,AFB_BaBar08}
shown in Fig.~\ref{fig:AFBexp}, and see that the current data is
already probing the difference between SM and the 4th generation
model, or the statement that Wilson coefficients $C_i$ could be
complex. As stated, SM expectation is ${\cal A}_{\rm FB} \sim
-0.03$ ({\it note the B factory sign convention}) for the region
$q^2 \in (4m_\mu^2$, 6.25 GeV$^2/c^4)$. This can be understood
from the solid line in Fig.~\ref{fig:AFB_NP}, where the
corresponding region is $\hat s < 0.22$. Since there is a crossing
over zero, and since the region below the zero is slightly larger
than above, we see that the SM expectation is slightly negative.
But Belle and BaBar data both indicate that ${\cal A}_{\rm FB} >
0$ is preferred, often phrased as $C_7 = -C_7^{\rm SM}$ seems
preferred from ${\cal A}_{\rm FB}$ data. This should be viewed as
just a way of expression, since it has been pointed
out~\cite{GHM05} that $C_7 = -C_7^{\rm SM}$, i.e. flipping the
sign of the photonic penguin, would lead to too large a $B\to
X_s\ell^+\ell^-$ rate as compared with experiment. It actually
illustrates our point to use Eq.~(\ref{eq:AFBcomplex}) rather than
Eq.~(\ref{eq:AFB}) in fitting data. In fact, we could even claim
that Belle and BaBar data favor somewhat the 4th generation curve
in Fig.~\ref{fig:AFB_NP}. Compared with the solid line, the zero
for the dashed line has moved to much lower $q^2$, together with a
drop in peak value. Therefore ${\cal A}_{\rm FB} > 0$ for the 4th
generation model motivated by $\Delta{\cal A}_{K\pi}$. Note that
this model predicts large and negative $\sin2\Phi_{B_s}$.

It is clear that the LHCb has good discovery potential using
${\cal A}_{\rm FB}$ to probe complexity of short distance Wilson
coefficients, without measuring CPV. In fact, {\it once again the
Tevatron could make earlier impact}. With 1 fb$^{-1}$ data, CDF
has demonstrated~\cite{rescigno06} branching ratio measurement
capability in $B^0\to K^{*0}\mu^+\mu^-$, comparable to that of
Belle and BaBar. Given that CDF and D$\emptyset$ expects to
accumulate of order 6 to 8 fb$^{-1}$ data per experiment, if such
studies could continue towards ${\cal A}_{\rm FB}$ measurement,
there is good potential for Tevatron to improve on Belle and BaBar
results, which would also be updated. A more definite statement on
whether SM is disfavored could come forth before LHCb data
arrives.

We note that if there are New Physics that affects the
$bs\ell\ell$ as a 4-quark operator, for example in $Z^\prime$
models with FCNC couplings, the allowed range for ${\cal A}_{\rm
FB}$ is practically unlimited. If such large effects are
uncovered, one would expect sizable direct CPV in $b\to
s\gamma$~\cite{HHM07}, which is another goal for Super B factory
studies.

\subsection{\boldmath $B\to K^{*}\nu\nu$}
\label{sec:Knunu}

The $B\to K^*\nu\nu$ (and $b\to s\nu\nu$) decay mode is attractive
from the theory point of view, since it can arise only from short
distance physics, such as $Z$ penguin and box diagram
contributions~\cite{HWS87}. The photonic penguin does not
contribute. In turn, these processes allow us to probe, in
principle, what happens in the loop. Interestingly, since the
neutrinos go undetected, the process also allows us to probe light
dark matter (DM), which is complementary to the DAMA/CDMS type of
direct search. This is because the latter type of experiments rely
on detecting special electronic signals arising from a nucleus
displaced by a DM particle. But this means that the approach loses
sensitivity for light DM particles. But for such particles, DM
pairs could arise from exotic Higgs couplings to the $b\to s$
loop.

BaBar has pioneered $B\to K^{*}\nu\nu$ search. More recently, as a
companion study to $B\to \tau\nu$ search, Belle has searched in
many modes with a large dataset of 535M $B\bar B$
pairs~\cite{Knunu_Belle07}, using the aforementioned method of
full reconstruction of the other $B$. No signal is found, and the
most stringent limit is $1.4 \times 10^{-5}$ in $B^+\to
K^+\nu\nu$. This is still a factor of 3 above the SM expectation
of $\sim 4\times 10^{-6}$ for this mode. However, it strengthens
the bound on light DM production in $b\to s$
transitions~\cite{Bird04}.
A complementary approach for search of light DM, as well as light
exotic Higgs bosons, is discussed in a different section.

%
It seems that, to measure the theoretically clean $B\to
K^{*}\nu\nu$ modes, one again requires a Super B factory.
Furthermore, here one really needs to improve on background
suppression, which seems challenging. After all, $B\to \tau\nu$
has just very recently been discovered through the technique of
fully reconstructing the other $B$, where the issues for improving
the measurements are common, i.e. the challenge of modes with
missing mass. Even with full reconstruction of the other $B$, one
probably needs to improve on detector hermeticity. We note that
there is no resort to LHCb for this mode. Thus, it should be an
emphasis for the Super B factory effort.

\section{\boldmath RH Currents and Scalar Interactions}
\label{sec:RHscalar}

It should be clear that loop-induced $b\to s$ transitions offer
many good probes of TeV scale New Physics. As last examples of
their usefulness, we discuss probing for right-handed (RH)
interactions via time-dependent $CP$ violation in $B^0\to
K_S^0\pi^0\gamma$ decay, and searching for enhancement of $B_s \to
\mu^+\mu^-$ as probe of BSM neutral Higgs boson effects. The
former is best done at a (Super) B factory, while the latter is
the domain of hadron colliders, where great strides have already
been made.

\subsection{\boldmath TCPV in $B\to X_0\gamma$}
\label{sec:SX0gamma}

With large QCD enhancement~\cite{bsgammaQCD}, the the $b\to
s\gamma$ rate is dominated by the SM. The left-handedness of the
weak interaction dictates that the $\gamma$ emitted in $\bar
B^0\to \bar K^{*0}\gamma$ decay has left-handed helicity (defined
somewhat loosely), with the emission of right-handed (RH) photons
suppressed by $\sim m_s/m_b$. This reflects the need for a mass
insertion for helicity flip, and the fact that a power of $m_b$ is
required by gauge invariance (or current conservation) for the
$b\to s\gamma$ vertex. For $B^0\to K^{*0}\gamma$ decay that
involves $\bar b\to \bar s\gamma$, the opposite is true, and the
emitted photon is dominantly of RH kind.

The fact that photon helicities do not match for $\bar B^0\to \bar
K^{*0}\gamma$ vs $B^0\to K^{*0}\gamma$ has consequences for a very
interesting probe~\cite{AGS97}. Mixing-dependent CPV, i.e. TCPV,
involves the interference of $\bar B^0$ and $\bar B^0
\stackrel{\rm mix}{\Longrightarrow} B^0$ decays to a common final
state that is not flavour-specific (i.e. no definite flavour). For
radiative $\bar B^0\to \bar K^{*0}\gamma$ decay (vs $\bar B^0
\stackrel{\rm mix}{\Longrightarrow} B^0\to K^{*0}\gamma$ decay),
the common final state is $K_S^0\pi^0$. Since the $\bar B^0\to
\bar K^{*0}\gamma$ process produces $\gamma_L$ while the $B^0\to
K^{*0}\gamma$ process gives $\gamma_R$, they cannot interfere as
they are orthogonal to each other! The interference is suppressed
by the helicity flip factor of $m_s/m_b \sim$ few \% within SM.
However, if there are RH interactions that also induce $b\to
s\gamma$ transition, then $\bar B^0\to \bar K^{*0}\gamma$ would
acquire a $\gamma_R$ component to interfere with the $\bar B^0
\Longrightarrow B^0\to K^{*0}\gamma$ amplitude. Thus, {\it TCPV in
$B^0\to K^{*0}\gamma$ decay mode probes RH interactions!}

Alas, Nature plays a trick on us. As mentioned, $K^{*0}\gamma$ has
to be in a $CP$ eigenstate, such as $K^{*0} \to K_S^0\pi^0$, so
the final state is $K_S^0\pi^0\gamma$. The $\pi^0$ and $\gamma$
certainly do not lead to vertices. For the $K_S$, though
``short-lived", it typically decays at the edge of the silicon
detector, and one has poor vertex information. Thus, it seems
impossible for TCPV to be studied in the $K_S^0\pi^0\gamma$ final
state, and the intriguing suggestion of Ref.~\cite{AGS97},
beautiful as it is, appeared to be just an impossible dream.
Fortunately, with a larger vertex detector than Belle with a extra
silicon plane, BaBar pushed forward a technique, called ``$K_S$
vertexing". It was demonstrated~\cite{SKspi0gam_BaBar04} that,
though degraded, the $K_S \to \pi^+\pi^-$ decay does give some
vertex information. The key point is the availability of the beam
direction information because of the boost, providing a ``beam
profile" for the somewhat rudimentary $K_S$ momentum vector to
point back to. The method was validated with gold plated modes
like $B^0\to J/\psi K_S$ (by removing the $J/\psi \to
\ell^+\ell^-$ tracks), and have been extended to TCPV studies such
as in $B^0\to K_SK_SK_S$.

The current status of TCPV in $B^0\to K^{*0}\gamma$ decay,
combining the 535M $B\bar B$ pair result from
Belle~\cite{Kstargamma_Belle06}, and the 232M result from
BaBar~\cite{Kstargamma_BaBar05}, gives the average of ${\cal
S}_{K_S\pi^0\gamma} = -0.28\pm 0.26$, which is consistent with
zero. A recent BaBar update with 431M
gives~\cite{Kstargamma_BaBar07} ${\cal S}_{K_S\pi^0\gamma} =
-0.08\pm 0.31\pm0.05$. Measurements have also been made in $B^0\to
K_s\pi^0\gamma$ mode without requiring the $K_s\pi^0$ to
reconstruct to a $K^{*0}$, as well as in the $B^0\to \eta
K_s\gamma$ mode.

This is a very interesting direction to explore, but again one
needs a Super B factory to seriously probe for RH interactions. At
the LHCb, which lacks the ``beam profile" technique for $K_S$
vertexing, the $B_s\to \phi\gamma$ mode may be used, although the
$\phi$ is not so good in providing a vertex, since the $K^+K^-$
pair is rather colinear because of $2m_K \sim m_\phi$. Probably
the LHCb upgrade would be needed to be competitive with a Super B
factory.
Other ideas to probe RH currents in $b\to s\gamma$ are $\gamma\to
e^+e^-$ conversion, $\Lambda$ polarization in $\Lambda_b \to
\Lambda\gamma$ decay, and angular $F_L$ and $A_T$ measurables in
$B\to K^*\ell^+\ell^-$.

\subsection{\boldmath $B_s\to \mu\mu$}
\label{sec:Bsmumu}

$B_s\to \mu^+\mu^-$ decay has been a favorite mode to probe exotic
Higgs sector effects in MSSM, because of possible large
$\tan\beta$ enhancement.

The process proceeds in SM just like $b\to s\ell^+\ell^-$, except
$\bar s$ is the spectator quark that annihilates the $b$ quark.
Since $B_s$ is a pseudoscalar, the photonic penguin does not
contribute, and one is sensitive to scalar operators. The SM
expectation is only $(3.4\pm 0.5)\times 10^{-9}$~\cite{Buras03},
because of $f_{B_s}$ and helicity suppression. In MSSM, a
$t$-$W$-$H^+$ loop can emit neutral Higgs bosons that turn into
muon pairs, giving rise to an amplitude $\propto
\tan^6\beta$~\cite{BK00}, which could greatly enhance the rate
even with modest pseudoscalar mass $m_A$. Together with the ease
for trigger and the enormous number of $B$ mesons produced, this
is the subject vigorously pursued at hadron facilities, where
there is enormous range for search.

With Run-II data now taking good shape, the Tevatron experiments
have improved the limits on this mode considerably. The recent 2
fb$^{-1}$ limits from CDF and D$\emptyset$ are $< 5.8 \times
10^{-8}$~\cite{BsmumuCDF08} and $9.3\times
10^{-8}$~\cite{BsmumuDzero07} respectively at 95\% C.L., combining
to give ${\cal B}(B_s\to \mu^+\mu^-) < 4.5 \times 10^{-8}$. This
is still an order of magnitude away from SM.

The expected reach for the Tevatron is about $2 \times 10^{-8}$.
Further improvement would have to come from LHCb. LHCb
claims~\cite{BsmumuLHCb} that, with just 0.05 fb$^{-1}$, it would
overtake the Tevatron, attain 3$\sigma$ evidence for SM signal
with 2 fb$^{-1}$, and 5$\sigma$ observation with 10 fb$^{-1}$. To
follow our suggested modest 0.5 fb$^{-1}$ expectation for the
first year of LHCb data taking, we expect LHCb to exclude
branching ratio values down to SM expectation.

Clearly, much progress will come with the turning on of LHC, where
direct search for Higgs particles and charginos would also be
vigorously pursued.

\section{\boldmath Bottomonium Decay and New Physics}
 \label{sec:UpsNP}

We make a detour from our $b\to s$ loop probes of New Physics, and
give some account of a special arena for New Physics search, in
the decays of bottomonium, namely $\Upsilon(nS)$, $n =1-3$. As we
have mentioned in Sec.~\ref{sec:Knunu}, the CDMS/DAMA type of
approaches for Dark Matter search are not sensitive to light DM.
The bottomonium offers to cover such a window. At the same time,
the related exotic Higgs sector can also be probed.

Motivated by a theoretical suggestion that ${\cal B}(\Upsilon(1S)
\to \chi\chi)$ could be of order 0.6\%~\cite{McElrath}, where
$\chi$ is a dark matter particle lighter than $m_b$, Belle made an
innovative special data run of 2.9 fb$^{-1}$ on the $\Upsilon(3S)$
to pursue DM search. Using $\pi^+\pi^-$ as kinematic tag for
$\Upsilon(1S) \to nothing$ in $\Upsilon(3S) \to
\pi^+\pi^-\Upsilon(1S)$ decay events, no signal was found, and a
limit below the theoretical prediction was
set~\cite{3SDM_Belle07}. This was followed by a search by
CLEO~\cite{1SDM_CLEO07} using 1.2 fb$^{-1}$ on the $\Upsilon(2S)$
for $\pi^+\pi^-\Upsilon(1S)$ decay where the $\Upsilon(1S)$ decays
invisibly. A limit slightly poorer than that of Belle is set.

When the PEP-II accelerator had to be terminated earlier than
scheduled because of US funding, Babar decided to take 30
fb$^{-1}$ on $\Upsilon(3S)$ (10 times Belle data) in early 2008,
followed by 15 fb$^{-1}$ on $\Upsilon(2S)$ (12 times CLEO data).
Further New Physics motivation for taking data on bottomonia came
from the potential to search for the exotic pseudoscalar Higgs
boson $a_1$ via $\Upsilon(1S) \to \gamma a_1$ followed by $a_1\to
\tau^+\tau^-$. The light $a_1$ could even be the 214.3 MeV
$\mu^+\mu^-$ events observed~\cite{Hyper05} by the HyperCP
experiment in $\Sigma^+\to p\mu^+\mu^-$, which provides further
motivation.

The DM search with 30 fb$^{-1}$ data on $\Upsilon(3S)$ awaits data
analysis. Let us elucidate the physics of light $a_1$ as follows.
In NMSSM (N stands for ``Next to"), a light pseudoscalar $a_1$
could be lighter than $2m_b$, allowing the SM-like neutral scalar
$H$ to be still lighter than 100 GeV, i.e. evade LEP bound, by
decaying via $H\to a_1a_1 \to 4\tau$. Such a scenario~\cite{DGM07}
is difficult to unravel at a hadronic collider. However, the light
$a_1$ can precisely be searched for in $\Upsilon \to \gamma a_1$
decay, where a lower bound on this rate is given~\cite{DGM07}.
If $a_1$ is lighter than $2m_\tau$, then $a_1 \to \mu^+\mu^-$
would dominate. It has been suggested that the 3 $\mu^+\mu^-$
events at 214.3 MeV as seen by HyperCP experiment at Fermilab,
could be~\cite{HTV07} such a light pseudoscalar.
Thus, besides DM search, BaBar was motivated to run on
$\Upsilon(3S)$ and $\Upsilon(2S)$ for direct radiative decay to
$a_1$, or via $\Upsilon(3S)$, $\Upsilon(2S) \to
\pi^+\pi^-\Upsilon$, followed by $\Upsilon \to \gamma a_1$.

Using 21.5M $\Upsilon(1S)$ collected by the CLEO III detector,
however, it was claimed~\cite{StoneFPCP08} very recently that most
of the parameter space for $2m_\tau < m_{a_1} < 7,5$ GeV, and {\it
all} the parameter space for light $a_1$ ($m_{a_1} < 2m_\tau$),
are ruled out.

It seems that, besides intrinsic interests in spectroscopy, the
bottomonium system also provides a window on New Physics. A future
Super B factory could probe this arena with ease, if flexible in
its energy reach.

\section{\boldmath D/K: Box and EWP Redux}
 \label{sec:D/K}

We touch upon $D$ and $K$ mesons only very briefly.

\subsection{\boldmath $D^0$ Mixing}
\label{sec:Dmix}

$D^0$-$\bar D^0$ mixing is the last neutral meson mixing to be
measured. Observation was claimed in 2007, which was quite some
feat of experimental effort.

Box diagrams, much like the $K^0$, $B_d^0$ and $B_s^0$ meson
systems, govern short distance contributions to $D^0$ mixing.
Unfortunately, the $d$ and $s$ quark masses are small compared to
$m_b$, hence only $b$ quark contributes in the box at short
distance. But $m_b$ is also tiny compared to $m_t$. Furthermore,
$V_{ub}V_{cb}^*$ is extremely small compared to the leading
$V_{ud}V_{cd}^*\simeq - V_{us}V_{cs}^* \cong -0.22$ in the CKM
triangle relation
\begin{eqnarray}
V_{ud}V_{cd}^*+V_{us}V_{cs}^*+V_{ub}V_{cb}^* = 0.
 \label{eq:ctouTri}
\end{eqnarray}
Thus, in the SM, $D^0$ mixing receives very tiny short distance
effects, making it susceptible to long distance contributions.

The quark level $s\bar s$ and $d\bar d$ intermediate states in the
box diagram are suppressed only by $V_{us}$ or $V_{cd}$, and
correspond to mesonic final states from $D^0$ decay. Common final
states for $D^0$ and $\bar D^0$ can cause interfere and generate a
{\it width difference}, much like in $K^0$-$\bar K^0$ and
$B_s^0$-$\bar B_s^0$ systems. It has been argued~\cite{Falk04}
that SU(3) breaking effects in $PP$ and $4P$ (where $P$ stands for
$K$ or $\pi$) final states can generate a percent level $y_D
\equiv \Delta\Gamma_D/2\Gamma_D$, the parameter usually used in
place of the width difference $\Delta\Gamma_D$. It was further
shown that a $y_D $ at the percent level can
generate~\cite{Falk04}, via a dispersion relation, width mixing
$x_D = \Delta m_D/\Gamma_D$ that is comparable in size to $y_D$.
Unfortunately, the hadronic uncertainties are uncontrollable. But
with the observation of $D^0$-$\bar D^0$ mixing in 2007, so far
$x_D \sim y_D \sim 1\%$ seems to be the case, i.e. consistent with
long distance effects.

\begin{figure}[t!]
\vskip0.3cm\hskip1.2cm
\includegraphics[width=0.35\textwidth,height=0.24\textwidth,angle=0]{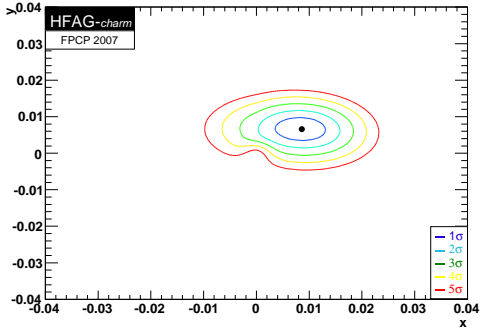}
 \vskip0.8cm
\caption{
 Observation of $D^0$ mixing: HFAG plot of combined fit to data,
 with Eq.~(\ref{eq:Dmixing}) as best fit result,
 together with $\delta_D = 0.33^{+0.26}_{-0.29}$ rad.}
 \label{fig:Dmixing}       
\end{figure}

The 2007 observation of $D^0$ mixing rests in i) Belle analysis of
540 fb$^{-1}$ data in $D^0\to K^+K^-$, $\pi^+\pi^-$ ($CP$
eigenstates) to extract $y_{CP}$~\cite{yCPBelle07}; ii) a Dalitz
analysis by Belle~\cite{D0Kspipi_Belle07} of $D^0\to
K_S\pi^+\pi^-$ with 540 fb$^{-1}$ to extract $x_D$ and $y_D$; iii)
both Belle and BaBar analyzed $D^0\to K^\mp\pi^\pm$ (Cabibbo
allowed vs suppressed), with 400 fb$^{-1}$ and 384 fb$^{-1}$ data
respectively, to extract $x_D^{\prime 2}$ and $y_D^\prime$, where
$x_D^{\prime}$ and $y_D^\prime$ is a rotation from $x_D$ and $y_D$
by a strong phase $\delta_D$ between the Cabibbo allowed and
suppressed $D^0\to K^\mp\pi^\pm$ decays. The analyses are too
complicated to report here. Suffice it to say that $(x_D,\; y_D) =
(0,\; 0)$ was excluded at the 5$\sigma$ level (see
Fig.~\ref{fig:Dmixing}), and $D^0$ mixing became established. The
best fit, assuming $CP$ invariance, gives,
\begin{eqnarray}
   x_D = 0.87^{+0.30}_{-0.34}\,\%, \ 
   y_D = 0.66^{+0.21}_{-0.20}\,\%, \ 
 \label{eq:Dmixing}
\end{eqnarray}
with $\delta_D = 0.33^{+0.26}_{-0.29}$ rad. While $y_D$ is more
solid, a finite \% level $x_D$ is indicated. Further progress has
been made after summer 2007. But rather than going into any
detail, we just quote the FPCP2008 results from HFAG~\cite{HFAG}.
As significance has been further improved, we quote the fit that
allows $CP$ violation (although data is consistent with no $CP$
violation),
\begin{eqnarray}
   x_D = 0.89^{+0.26}_{-0.27}\,\%, \ 
   y_D = 0.75^{+0.17}_{-0.18}\,\%, \;\ ({\rm FPCP2008}) 
 \label{eq:Dmixing08}
\end{eqnarray}
with $\delta_D = (21.9^{+11.3}_{-12.4})^\circ$.

It is of interest to note that, if the $4P$ final state dominates
the long distance contribution, which is consistent with $y_D \sim
1\%$, then $x_D^{\rm LD}$ and $y_D$ (necessarily long distance)
should be of the opposite sign~\cite{Falk02}, while data show the
same sign. Although it has been checked~\cite{Falk04} that
changing hadronic parameters does not change this conclusion,
unfortunately the hadronic effects are not well under control to
make a definite statement.
In any case, one should remember the $\Delta m_K$ enterprise of
20-30 years ago. That is, although the observed strength could
arise from long distance effects, comparable BSM, at twice the
observed $x_D$, is always allowed.

Besides continued progress, there are two things to watch in
regards $D^0$ mixing. While other measurements have seen steady
progress for several years, it is for the first time that the
Dalitz analysis of Belle~\cite{D0Kspipi_Belle07} sees an
indication for $x_D$. Second, to unravel some of the hadronic
physics in the decay final state, one needs to gain access to
strong phases. By a tagged Dalitz analysis in $\psi(3770) \to
D^0\bar D^0$, one can~\cite{AS06} extract the strong phase
$\delta_D$, which would in turn feedback on $x$ and $y$
extraction. Unfortunately, CLEO-c ended up not taking enough data
on the $\psi(3770)$ resonance before shutdown. But in the future,
BES-III and other possible charm factories could aid the $D^0$
mixing program considerably through this type of studies.
Basically, the Dalitz type of analysis, with the help of quantum
coherence, holds the power for the future.

What we are interested in is the New Physics impact, rather than
hadronic physics. For the moment, though one has made great
experimental stride, there is no indication of New Physics in
$D^0$ mixing. A comprehensive study for New Physics implications
can be found in Ref.~\cite{GHPP07}. Ultimately it seems, one would
need to measure CPV, expected to be tiny within SM (with or
without long distance dominance), to find unequivocal evidence for
BSM.
This is an area where a Super B factory can compete well with LHCb
because of its diversity. However, LHCb can also play a role, as
evidenced by the CDF study~\cite{DmixCDF08} of $D^0\to
K^\pm\pi^\mp$ mode with 1.5 fb$^{-1}$, which has results that are
complementary to Belle and BaBar in this mode.

\subsection{\boldmath Rare $K$ Decays}
\label{sec:rareK}

This field saw its last hurrah in $\varepsilon^\prime/\varepsilon$
almost a decade ago. Despite the top effect through the
electroweak penguin allowed it to vanish, unfortunately, the
interpretation of $\varepsilon^\prime/\varepsilon$ is almost
completely clouded by long-distance effects.

With the cancellations of CKM at Fermilab and KOPIO at BNL, the
kaon program in the US has withered,\footnote{
 Can the US revamp its kaon program with Project-X at Fermilab?
 Let's wait and see.
 }
despite a long standing hint of 3 events for $K^+\to \pi^+\nu\nu$
at BNL by E787/949. At CERN, there is the P236 proposal (has
become NA62) to use the SPS, aiming at reaching ${\cal O}(80)$
events with 2 years of running, assuming the SM branching ratio of
$\sim 10^{-10}$. Once approved, data taking could start in 2012.
If successful, the hope is to upgrade the CERN proton complex
towards ``EUREKA" (European Rare-decays Experiments with Kaons).

In Japan, one has the E391A experiment at KEK PS, which just came
out with a new limit~\cite{E391A} on $K_L\to \pi^0\nu\nu$, of less
than $6.6\times 10^{-8}$ at 90\% C.L., improving its previous
limit by a factor of 3. Another dataset equivalent in size is
being analyzed. Though one is far from probing the SM expectation
of $10^{-11}$, there is New Physics potential. But E391A should be
viewed as the pilot study for the more ambitious E14 proposal to
the J-PARC facility, which aims at eventually reaching below
$10^{-12}$ sensitivity to probe BSM. The first step for E14,
besides a new beam-line, is to use the upgraded E391A detector
(e.g. with CsI crystals from the KTeV experiment at Fermilab). The
earliest start is 2011, hopefully seeing 1 event with SM branching
ratio. If there is New Physics enhancement, then discovery could
come earlier, but if SM persists, then a 10\% measurement
requiring ${\cal O}(100)$ events, and it would probably take a
decade from the present time.

The $K^+\to \pi^+\nu\nu$ and $K_L\to \pi^0\nu\nu$ decays are clean
modes theoretically, and especially the latter holds big room for
discovering BSM physics. The challenge is to get the experiment
done, but these are some years away.

\section{\boldmath $\tau$: LFV and $(B-L)$V}
\label{sec:tau}

Before concluding, we touch upon exciting developments in rare tau
decays: radiative decays which have $b\to s$ echoes, and the
enigmatic (if found) baryon number violating decays. There should
be no doubt that we would have uncovered Beyond the Standard Model
physics if any of these are observed.

\subsection{\boldmath $\tau\to \ell\gamma$, $\ell\ell\ell^\prime$}
\label{sec:tau_radiative}

The $\tau\to \ell\gamma$ processes are extremely suppressed in SM
by the very light neutrino mass. This opens up the opportunity to
probe BSM, just like the venerable $\mu\to e\gamma$ (where there
is the fabulous MEG experiment at PSI). Observation of lepton
flavor violating (LFV) decays would definitely mean New Physics!
Besides, there is also the backdrop of large neutrino mixings.
Again, the favorite is SUSY, ranging from sneutrino-chargino
loops, exotic Higgs, $R$-parity violation, $\nu_R$ in SO(10), or
large extra dimensions (LED). Predictions for $\tau\to\mu\gamma$,
$\ell\ell\ell$, $\ell\ell\ell^\prime$, $\ell M^0$ (where $M^0$ is
a neutral meson) could reach the $10^{-7}$ level. The models are
often well motivated from observed near maximal
$\nu_\mu$-$\nu_\tau$ mixing, or interesting ideas such as
baryogenesis through leptogenesis. The great progress in neutrino
physics of the past decade has stimulated a lot of interest in
these LFV decays.

On the experimental side, the stars are once again the B
factories: With $\sigma_{\tau\tau} \sim 0.9\,$nb comparable to
$\sigma_{bb} \sim 1.1\,$nb, B factories are also $\tau$ (and
charm) factories! As data increased steadily, the B factories have
pushed the limits from $10^{-6}$ of the CLEO era, down to the
$10^{-8}$level. For example, with the 535 fb$^{-1}$ analysis by
Belle~\cite{tauto3l_Belle07} the limits on
$\tau\to\ell\ell\ell^\prime$ modes such as $\mu^+e^-e^-$ and
$e^+\mu^-\mu^-$ have reached $2\times 10^{-8}$, with BaBar not far
behind~\cite{tauto3l_BaBar07}. Thus, some models or in their
parameter space are now ruled out.
%

With BaBar closed, and with Belle at best giving result at 1
ab$^{-1}$, one at best touches the $10^{-8}$ boundary. To probe
deeper into the parameter space of various LFV rare $\tau$ decays,
a Super B factory would be called for. In the near future, LHCb
can compete in the all charged track modes, but modes with
neutrals would be difficult.

\subsection{\boldmath $\tau\to \Lambda\pi$, $p\pi^0$}
\label{sec:tau_BNV}

A somewhat wild idea is to search for baryon number violation
(BNV) in $\tau$ decay, i.e. involving the 3rd generation. This was
pointed out in Ref.~\cite{BNM05}, but the same reference argued
that, by linking to the extremely stringent limit on proton decay,
BNV ($B-L$ violating to be more precise) involving higher
generations are in general much too small to be observed. This did
not stop Belle from conducting a search~\cite{BNM_Belle05},
followed by BaBar~\cite{BNM_BaBar06}. So far, no signal is found,
as expected.

\section{Discussion and Conclusion}
\label{sec:Conclusion}

The last subsection brings us to wilder speculations that we have
shunned so far. In the SUSY conference, however, ideas range
widely, if not wildly. To this author, from an experimental point
of view, the question is identifying the smoking gun, or else it
is better to stick to the simplest explanation of an effect that
requires New Physics. That has been our guiding principle.

Perhaps the wildest idea in 2007, and probably the one bringing
out the most insight, is ``unparticle physics"~\cite{UnP07}. We do
not discuss what this is all about, but it has clearly stimulated
much (theoretical) interest. On the flavour and CPV front, for
example, there is the suggestion that unparticles could generate
DCPV in unexpected places~\cite{dcpvUnP07}. Sure enough, this
suggestion may well have been stimulated by the 3.2$\sigma$
indication~\cite{dcpvD+D-_Belle07} of DCPV in $B^0\to D^-D^+$ by
Belle (though the BaBar result is consistent with
zero~\cite{dcpvD+D-_BaBar07}) that is otherwise very difficult to
explain. We note also that further studies of other $B^0 \to
D^{(*)}\bar D^{(*)}$ modes have {\it not} revealed anything to
support the evidence for DCPV in $B^0\to D^-D^+$. So, the Belle
result needs to be revisited with more data. But searching for
DCPV in the $B^+\to \tau^+\nu$ mode is also
suggested~\cite{dcpvUnP07}, which is interesting. If I may
speculate, maybe unparticles could generate BNV in the modes of
the previous subsection. In any case, new ideas such as these
stimulate search efforts in otherwise unmotivated places, hence
are very valuable.

To summarize, I have covered a rather wide range of probes of TeV
scale physics via heavy flavour processes. At the moment, we have
two hints for New Physics: in the $\Delta {\cal S}$ difference
between TCPV in $B\to J/\psi K^0$ vs penguin dominant $b\to s\bar
qq$ modes; and in the experimentally established difference in
DCPV between $B^+\to K^+\pi^0$ and $B^0\to K^+\pi^-$ modes. These
are large CPV effects, but they are not unequivocal, either in
experimentation, or in interpretation. Because of this, the thing
to watch in 2008-2009, in my opinion, is whether the Tevatron
could see a hint for {\it large} mixing-dependent CPV in $B_s\to
J/\psi\phi$, i.e. $\sin2\Phi_{B_s}$. If seen, it would be
unequivocal as evidence for BSM. Curiously, a hint has appeared by
Winter 2008. Though still too early to conclude, it should be
clear that Tevatron can have 3-4 times the data than analyzed, and
{\it the hint could turn into evidence, before LHCb physics
arrives}. In any case, if the hint for sizable $\sin2\Phi_{B_s}$
is true, it can be quickly confirmed by LHCb. If the hint for
$\sin2\Phi_{B_s}$ Tevatron fades away, LHCb can probe down to SM
expectation rather quickly, with still a lot of range for New
Physics discovery. But it would be a great disappointment if we
again confirm the Standard Model. Other processes that have good
potential for New Physics search emphasized in this brief review
are: direct CPV in $B^+\to J/\psi K^+$; $B\to\tau\nu$; $b\to
s\gamma$; ${\cal A}_{FB}(B\to K^*\ell^+\ell^-$); $B_s\to \mu\mu$;
$\Upsilon$ decay; $D^0$ mass mixing and CPV; and $\tau \to
\ell\gamma$, $\ell\ell\ell^\prime$.

Though no unequivocal indication for New Physics has emerged so
far, the B factories have not yet exhausted their bag of
surprises. With such a diverse search platform, I hope I have made
it clear that a Super B factory would be superb to probe deeper
into all the above subjects (except maybe $B_s\to \mu\mu$). Before
that, we will attain some new heights with LHCb.

\vskip0.7cm

\noindent{\bf Epilogue: A tribute to Julius Wess}

\vskip0.3cm

The worldlines of Julius Wess' and mine have never really crossed.
Even at SUSY 2007, I watched him only at a distance (and was
shocked and saddened to learn his sudden death shortly
thereafter). I was in Munich for two years, at the
Werner-Heisenberg-Institut, but I left the year before he moved
from Karlsruhe to become a director at the Max-Planck-Institut
f\"ur Physik. But of course I remember my (theory) graduate
student days in the early 1980's, where, it seems, if you don't
learn the latest on SUSY, you're a goner. Julius Wess loomed as a
demigod. Then it was the second superstring revolution, and by
1986, I was gone for good, into phenomenology, venturing into B
physics and CP violation, even becoming a (pseudo)experimentalist
by the mid-1990s.

Of course, I have no doubt about the contribution and impact that
Julius Wess has made. I agree with his ``personal belief" that a
symmetry as beautiful as supersymmetry may be much more
fundamental, and would play a role at higher energies~\cite{Wess}.
The hierarchy problem may be too small, if not artificial, for
SUSY, as it is GUT inspired. To me, the natural scale for SUSY
ought to be the Planck scale, since naively speaking, two SUSY
transforms make a spacetime transform, while we know gravity holds
a fundamental scale, the Newton constant.

From the perspective of the task given me at SUSY 2007, the
experimental view on the link of Flavour and TeV scale physics, I
would offer the thankless remark that Flavour and SUSY are
``orthogonal". Low energy SUSY (rather than the idealized
beautiful symmetry as beheld by Julius Wess) is one of the oldest
tools in our arsenal to stabilize the electroweak symmetry
breaking scale. But in the parallel symmetry table of spacetime
symmetry vs inner space by Julius Wess, there is no mentioning of
Flavour. Thus, this was not his main concern. The symmetry, if
any, behind Flavour is not quite understood. In fact, it poses
some embarrassment for SUSY when considering Flavour: If SUSY is
broken, why don't we have FCNC all over the place? Thus, for the
SUSY practitioner, Flavour violation is banished into a corner.
Even so, Flavour physics haunts, and clamps down on, the SUSY
parameter space. I would confess that, besides chasing the
bandwagon in my student days, for a long time I prided myself in
having never written a paper on SUSY --- until the arrival of the
B factories, whence I wrote on SUSY in the Flavour context. That
is still a tribute to the impact on me by Julius Wess.

While we are at the brink of major progress in probing into
symmetry breaking physics, from the experimental side, Flavour
physics has its own, complementary life. I would stress that all
known effects of CPV rest in Yukawa couplings, which appear to be
dynamical couplings that have not yet found a Symmetry principle
foundation. For that matter, I offer the answer to why I have so
often emphasized the possibility of a fourth sequential generation
in this contribution on Flavour-TeV link: the 4 generation
``Standard Model" can enhance the traditionally held Jarlskog
invariant of 3 generations, the venerable $10^{-20}$, by 15 orders
of magnitude~\cite{CPVBAUQ}, thereby providing enough CPV for
baryogenesis. It is about large Yukawa coupling enhancement, which
we already see in the top quark. Considering the Baryon Asymmetry
of the Universe, maybe there really is a 4th generation. And that
may change our attitude on SUSY.

\vskip0.7cm

\noindent{\large \bf \textsf{Appendix: A {\boldmath $CP$}
Violation Primer}}

\vskip0.3cm

CPV is defined as a difference in probability between a particle
process from the antiparticle process, e.g. between $B\to f$ and
$\bar B \to \bar f$. It requires the presence of two interfering
amplitudes. But besides the usual $i$ from quantum mechanics, it
needs {\it complex dynamics} as well. That is, the interference
involves the presence of two kinds of phases. Let us elucidate how
CPV occurs.

Consider the particle process amplitude $A = A_1 + A_2$, which is
a sum of two terms, where amplitude $A_j$ has both a $CP$
invariant phase $\delta_j$ (imaginary $i$ from QM) and a CPV phase
$\phi_j$ (imaginary $i$ from CPV dynamics). Absorbing an overall
phase by defining $A_1 = a_1$ to be real, one has
\begin{eqnarray}
A &=& A_1 + A_2 = a_1 + a_2 e^{i\delta}e^{+i\phi},\nonumber\\
\bar A &=& \bar A_1 + \bar A_2 = a_1 + a_2 e^{i\delta}e^{-i\phi},
 \label{eq:A1plusA2}
\end{eqnarray}
where $a_2 \equiv \vert A_2\vert$. The $\delta$ and $\phi$ are
called the ``strong" and weak phases, respectively. The QM or
strong phase $\delta$ does not distinguish between particle or
antiparticle, hence sign is unchanged. However, the dynamical or
weak phase $\phi$ changes sign for the antiparticle process $\bar
A$. This enrichment of quantum interference leads to an asymmetry
between particle and antiparticle probabilities,
\begin{eqnarray}
{\cal A}_{\rm CP} &\equiv&
        \frac{\Gamma_{\bar B^0 \to \bar f} - \Gamma_{B^0 \to f}}
             {\Gamma_{\bar B^0 \to \bar f} + \Gamma_{B^0 \to f}}
                   \nonumber\\       &=&
        \frac{2a_1a_2 \sin{\delta}\sin{\phi}}
             {a_1^2+a_2^2 + 2a_1a_2 \cos{\delta}\cos{\phi}},
 \label{eq:ACP}
\end{eqnarray}
defined with respect to quarks. As ${\cal A}_{\rm CP}$ vanishes
with either $\delta$ or $\phi \to 0$, CPV requires the presence of
both $CP$ conserving {\it and} CPV phases.

\begin{figure}[t]
\centering
\vskip0.3cm
\includegraphics[height=2.5cm]{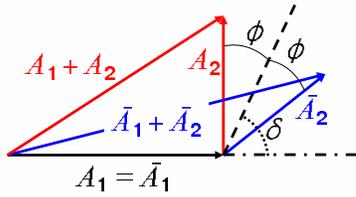}
\vskip0.2cm
%
%
\caption{Mechanism for CPV, Eq.~(\ref{eq:ACP}).}
\label{fig:CPV}       
\end{figure}

Eq.~(\ref{eq:A1plusA2}) is illustrated in Fig.~\ref{fig:CPV},
which shows geometrically how Eq.~(\ref{eq:ACP}) materializes. If
$\delta = 0$, then $A_1+A_2$ and $\bar A_1 + \bar A_2$ are at
angle $\phi$ above or below the real axis, hence of equal length.
If $\phi=0$, then $A_1+A_2$ and $\bar A_1 + \bar A_2$ are the same
vector. Only when {\it $\delta \neq 0$ and $\phi \neq 0$} does
$|A_1+A_2| \neq |\bar A_1 + \bar A_2|$ occur, which gives the
asymmetry of Eq.~(\ref{eq:ACP}).

In the KM model with 3 generations, the CPV phase is put in the 13
and 31 elements ($V_{ub}$ and $V_{td}$) in the standard phase
convention~\cite{PDG}. Thus, it is said that one needs the
presence of all 3 generations to make CPV to occur.

%
%

\end{document}